\documentclass[prd,twocolumn,nopacs,floatfix,amsmath,nofootinbib,amssymb,preprintnumbers]{revtex4-2}
\usepackage{graphicx,subfigure,color,dcolumn,booktabs}
\usepackage{array}
\usepackage{tabularx}
\usepackage{txfonts}
\usepackage{slashed}
\usepackage{epstopdf}
\usepackage{amsmath}
\usepackage{appendix}
\usepackage[colorlinks,
citecolor=blue,
anchorcolor=red,
menucolor=red,
linkcolor=red,
filecolor=red,
runcolor=red,
urlcolor=blue,
frenchlinks=red]{hyperref}
\usepackage{bm}
\graphicspath{{Figures/}}

\begin{document}
\newcommand{\pdlr}{\overset{\leftrightarrow}{\partial}}
\newcommand{\vect}[1]{\boldsymbol{#1}}

\title{Discovery potential of charmonium $2P$ states through the $e^+e^- \to \gamma D\bar{D}$ processes}
\author{Tian-Le Gao$^{1,2,4}$}
\author{Ri-Qing Qian$^{1,2,4}$}
\author{Xiang Liu$^{1,2,3,4}$}\email{xiangliu@lzu.edu.cn}
\affiliation{
	$^1$School of Physical Science and Technology, Lanzhou University, Lanzhou 730000, China\\
	$^2$Lanzhou Center for Theoretical Physics, Key Laboratory of Theoretical Physics of Gansu Province, Key Laboratory of Quantum Theory and Applications of MoE and Gansu Provincial Research Center for Basic Disciplines of Quantum Physics, Lanzhou University, Lanzhou 730000, China\\
	$^3$MoE Frontiers Science Center for Rare Isotopes, Lanzhou University, Lanzhou 730000, China\\
	$^4$Research Center for Hadron and CSR Physics, Lanzhou University and Institute of Modern Physics of CAS, Lanzhou 730000, China
}

\begin{abstract}
	In this work, we investigate the production of charmonium $2P$ states via the $e^+e^-\to \gamma D\bar{D}$ process at $\sqrt{s} = 4.23$ GeV. Using the measured cross-section data for $e^+e^-\to \gamma X(3872)$ as a reference, we calculate the cross sections for $e^+e^-\to \gamma \chi_{c0}(2P)$ and $e^+e^-\to \gamma \chi_{c2}(2P)$. Since the $\chi_{c0}(2P)$ and $\chi_{c2}(2P)$ states predominantly decay into $D\bar{D}$ final states, we also predict the corresponding $D\bar{D}$ invariant mass spectrum for the $e^+e^-\to \gamma D\bar{D}$ process. Our results indicate that $e^+e^-\to \gamma D\bar{D}$ is an ideal process for identifying the $\chi_{c0}(2P)$ and $\chi_{c2}(2P)$ states, analogous to the $\gamma\gamma\to D\bar{D}$ and $B^+\to D^+D^-K^+$ processes. This study highlights the discovery potential of charmonium $2P$ states at BESIII and Belle II.
\end{abstract}

\maketitle

\section{introduction}

Although 50 years have passed since the discovery of the $J/\psi$ particle, the first identified charmonium state, the construction of the charmonium family remains an ongoing challenge in the study of hadron spectroscopy. 
A typical example is the identification of $2P$ triplet $\chi_{cJ}(2P)$ ($J=0,1,2$) of charmonium.

In 2003, the charmonium-like state $X(3872)$ was observed in the $B \rightarrow J/\psi \pi^+ \pi^- K$ decay~\cite{Belle:2003nnu}. Its mass was found to be lower than the value predicted for the charmonium state $\chi_{c1}(2P)$ by the quenched potential model~\cite{Barnes:2005pb}, raising questions about its nature. To address this mass discrepancy, the $X(3872)$ was proposed to be an exotic hadronic state, such as a compact tetraquark~\cite{Maiani:2004vq,Hogaasen:2005jv,Cui:2006mp} or a $D\bar{D}^*$ molecular state~\cite{Wong:2003xk,AlFiky:2005jd,Wang:2019mhs}. However, with advancements in unquenched potential models, the low-mass puzzle no longer precludes the classification of the $X(3872)$ as part of the charmonium family. 

In fact, the charmonium $2P$ states form a spin triplet: in addition to the $\chi_{c1}(2P)$, there are the $\chi_{c0}(2P)$ and $\chi_{c2}(2P)$. Subsequently, the Belle Collaboration discovered a charmoniumlike state, the $Z(3930)$, in the $D\bar{D}$ invariant mass spectrum via the $\gamma\gamma \to D\bar{D}$ process~\cite{Belle:2005rte}. This state was convincingly identified as $\chi_{c2}(2P)$~\cite{Liu:2009fe}. However, the identification of $\chi_{c0}(2P)$ has been filled with challenges and complexities.
In 2009, the Belle Collaboration reported the observation of a charmonium-like state, the $X(3915)$, in the $\gamma\gamma \to J/\psi\omega$ process. According to the Lanzhou group's analysis, the $X(3915)$ was a strong candidate for the charmonium $\chi_{c0}(2P)$ state, and it was subsequently included in the charmonium section of the Particle Data Group (PDG) in the 2012 edition of the Review of Particle Physics (RPP)~\cite{ParticleDataGroup:2012pjm}. However, this assignment raised several important questions~\cite{Guo:2012tv,Olsen:2014maa}: Why is the mass gap between the $X(3915)$ ($\chi_{c0}(2P)$) and $Z(3930)$ ($\chi_{c2}(2P)$) so small? Why is there no observed signal for the $X(3915)$ in the $D\bar{D}$ invariant mass spectrum from $\gamma\gamma \to D\bar{D}$ process?
To address these issues, the Lanzhou group conducted further studies~\cite{Chen:2012wy,Duan:2020tsx}. They demonstrated that the small mass gap could be well reproduced by incorporating unquenched effects into the charmonium spectrum. Additionally, they suggested that the observed 3.9 GeV enhancement in the $D\bar{D}$ invariant mass spectrum from $\gamma\gamma \to D\bar{D}$ might comprise two substructures, corresponding to the $\chi_{c0}(2P)$ and $\chi_{c2}(2P)$~\cite{Chen:2012wy}.

Finally, in 2020, the LHCb Collaboration observed both the $\chi_{c0}(2P)$ and $\chi_{c2}(2P)$ in the $D^+D^-$ invariant mass spectrum via the $B \to K D\bar{D}$ process~\cite{LHCb:2020pxc,LHCb:2020bls}. This observation confirmed the predicted small mass gap between the $\chi_{c0}(2P)$ and $\chi_{c2}(2P)$~\cite{Duan:2020tsx}. The establishment of the $X(3915)$ as the $\chi_{c0}(2P)$ was ultimately achieved through the combined efforts of both theorists and experimentalists \cite{Belle:2005rte,Belle:2009and,Liu:2009fe,Chen:2013yxa,Chen:2012wy,Duan:2020tsx,Duan:2021bna,LHCb:2020pxc,LHCb:2020bls}\footnote{The authors of Ref. \cite{Ji:2022vdj} analyzed the relevant data and claimed alternative assignments for the $X(3915)$, which would need to be further examined using high-precision data.}.

As mentioned above, the $D\bar{D}$ channel plays a crucial role in identifying the properties of the $\chi_{c0}(2P)$ and $\chi_{c2}(2P)$. In addition to the $B \to K D\bar{D}$ decay, we aim to explore other processes that may produce the $D\bar{D}$ invariant mass spectrum. Motivated by this, we propose in this work to investigate the charmonium $2P$ states through the process $e^+e^- \to \gamma D\bar{D}$, which is closely related to several experimental observations.

In 2014, the BESIII Collaboration analyzed the $e^+e^- \to \gamma X(3872)$ process and observed a resonance-like structure around 4.2 GeV in the $\gamma X(3872)$ invariant mass spectrum~\cite{BESIII:2013fnz}. This marked the first observation of $X(3872)$ produced via radiative decay. With the accumulation of additional experimental data, the BESIII Collaboration revisited this process in 2019~\cite{BESIII:2019qvy}. The measured cross section for the process $e^+e^- \to \gamma X(3872)$~\cite{BESIII:2013fnz,BESIII:2019qvy} provides a valuable reference for investigating similar processes, such as $e^+e^- \to \gamma \chi_{c0,2}(2P) \to \gamma D\bar{D}$.

In this work, we investigate the $e^+e^- \to \gamma \chi_{cJ}(2P)$ processes. Observing that the resonance structure near 4.2 GeV is close to the charmonium state $\psi(4230)$ \cite{He:2014xna,Li:2009zu,Chen:2017uof,Wang:2019mhs}, we assume that these processes proceed via the intermediate state $\psi(4230)$. The partial widths of the $\psi(4230) \to \gamma \chi_{cJ}(2P)$ transitions are calculated using a hadronic loop model involving charmed mesons, which effectively captures the coupled-channel effects of excited charmonium states.

Using the calculated ratios of partial widths $\Gamma_{\gamma\chi_{c0}}:\Gamma_{\gamma\chi_{c1}}:\Gamma_{\gamma\chi_{c2}}$, combined with available experimental data, we predict the production cross sections for $e^+e^- \to \gamma \chi_{c0,2}(2P)$. Additionally, we focus on the $e^+e^- \to \gamma D\bar{D}$ process, providing predictions for the $D\bar{D}$ invariant mass spectrum, which can serve as a valuable guide for experimental studies.
Given the small mass gap between the $\chi_{c0}(2P)$ and $\chi_{c2}(2P)$, we also explore the angular distributions of the $\psi(4230) \to \gamma \chi_{c0}(2P)$ process. This analysis aims to identify distinctive features that could help differentiate between these two states.

This paper is organized as follows. After the introduction, we analyze the $e^+e^- \rightarrow \gamma \chi_{cJ}(2P)$ processes and calculate the partial widths of $\psi \to \gamma \chi_{cJ}(2P)$ using the hadronic loop mechanism in Sec.~\ref{sec2}. In Sec.~\ref{sec3}, we present the cross sections of the $e^+e^- \rightarrow \gamma \chi_{cJ}(2P)$ processes and further examine the contributions of the $\chi_{c0}(2P)$ and $\chi_{c2}(2P)$ states to the $D\bar{D}$ invariant mass spectrum in the $e^+e^- \rightarrow \gamma \chi_{c0,2}(2P) \rightarrow \gamma D\bar{D}$ processes. Finally, the paper concludes with a brief summary.

\section{The production of the $\chi_{cJ}(2P)$ by $e^+e^-\to \gamma D\bar{D}$}\label{sec2}

    In light of the observation of both $\chi_{c0}(2P)$ and $\chi_{c2}(2P)$ states in the $D^+D^-$ invariant mass spectrum by the LHCb Collaboration~\cite{LHCb:2020pxc,LHCb:2020bls}, it is crucial to investigate the $D\bar{D}$ invariant mass spectrum of these states through additional production mechanisms. The BESIII Collaboration's measurement of the cross section for $e^+e^-\to\gamma X(3872)$ \cite{BESIII:2013fnz,BESIII:2019qvy} reveals that the accumulation of data around 4.2 GeV is likely due to the $\psi(4230)$\footnote{Theoretical predictions for the search of $e^+e^-\to\gamma X(3872)$ at $\sqrt{s}=4.26$ GeV were made in Refs. \cite{Guo:2013zbw}, based on the interpretation of the observed charmonium-like structure $Y(4260)$ as a $D_1\bar{D}$ molecular state \cite{Ding:2008gr,Li:2013bca,Wang:2013cya}, while the $X(3872)$ is considered as a $D\bar{D}^*$ molecule. However, in 2017, high-precision data of $e^+e^-\to J/\psi\pi^+\pi^-$ \cite{BESIII:2016bnd} led to the replacement of the $Y(4260)$ by two substructures, $Y(4230)$ and $Y(4330)$. The $Y(4230)$ serves as a scaling point for constructing higher charmonium spectroscopy above 4 GeV \cite{Wang:2019mhs,Wang:2020prx} when assigning the $Y(4230)$ as a $4S$-$3D$ mixed state within the charmonium family. As demonstrated in coupled-channel analysis \cite{Lu:2017yhl,Man:2025zfu}, the contribution of the $D_1(2420) \bar{D}$ channel is significant, comparable in magnitude to those from the $D^{(*)}\bar{D}^{(*)}$ channels~\cite{Man:2025zfu}. Consequently, a realistic theoretical framework must account for both hadronic loops involving $S$-wave charmed mesons and the explicit inclusion of the $D_1(2420) \bar{D}$ channel. }, which has been extensively studied and can be consistently interpreted as a charmonium state in an unquenched picture \cite{He:2014xna,Li:2009zu,Chen:2017uof,Wang:2019mhs}. Therefore, the $e^+e^-\rightarrow\gamma X(3872)$ process can be viewed as a sequential process $e^+e^-\to\psi \to \gamma X(3872)$. The production of $P$-wave heavy quarkonia from higher $S$-wave heavy quarkonia radiative decays is well-established in similar processes, such as $\psi(2S) \to \gamma \chi_{cJ}$ \cite{BES:2005bmx}, $\Upsilon(2S) \to \gamma \chi_{bJ}$ \cite{CLEO:2004jkt}, and $\Upsilon(3S) \to \gamma \chi_{bJ}(2P)$~\cite{CLEO:2004jkt}.
Given this, if we treat the $X(3872)$ as the charmonium state $\chi_{c1}(2P)$, analogous processes like $e^+e^-\to \psi \to \gamma \chi_{c0,2}(2P) \to \gamma D\bar{D}$ could provide a promising path for the simultaneous observation of the $\chi_{c0}(2P)$ and $\chi_{c2}(2P)$ states. Notably, evidence of structures above 3.9 GeV has been observed in the $\omega J/\psi$ invariant mass spectrum in the process $e^+e^-\to\gamma J/\psi \omega$~\cite{BESIII:2019qvy}, which may further support this investigation.

\begin{figure}[htbp]
		\centering
		\includegraphics[width=8.6cm]{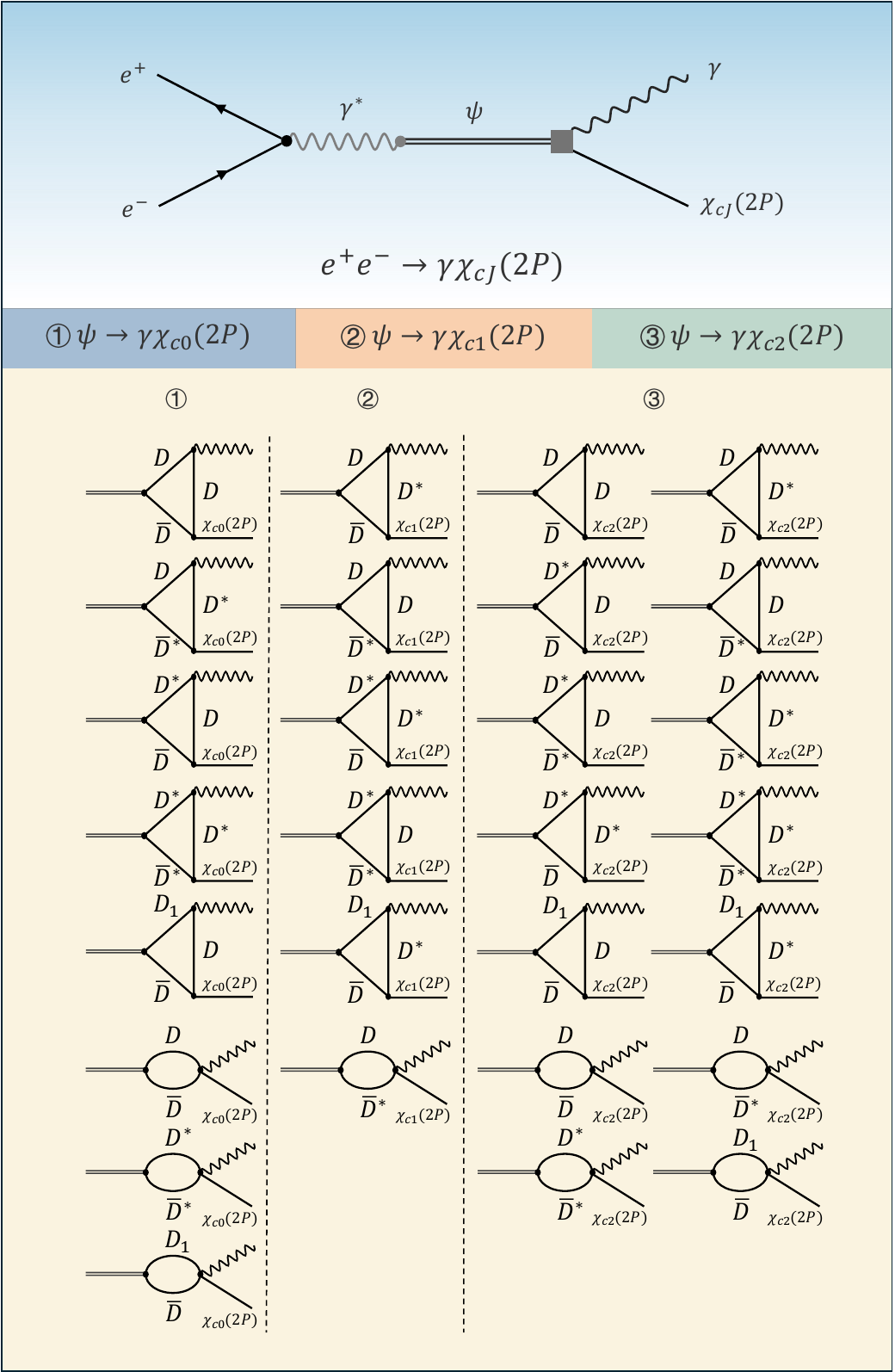}
		\caption{Schematic diagram of the $e^+e^- \to \gamma \chi_{cJ}(2P)$ processes and the corresponding decay diagrams for $\psi \to \gamma \chi_{cJ}(2P)$, incorporating the hadronic loop mechanism.}
		\label{diagram}
    \end{figure}

    Based on the above analysis, the production of $\gamma \chi_{cJ}(2P)$ in $e^+e^-$ annihilation is illustrated in Fig.~\ref{diagram}. The cross sections for $e^+e^-\to\gamma \chi_{cJ}(2P)$ can be expressed as
    \begin{align}
		  \sigma(e^+e^-\rightarrow \psi\rightarrow\gamma\chi_{cJ}(2P))
		=  \frac{12\pi\Gamma^{e^+e^-}_\psi\Gamma_{\gamma\chi_{cJ}}}{|s-m^2_\psi+im_\psi\Gamma_\psi|^2},
		\label{sigma}
	\end{align}
where $s$ is the square of the center-of-mass energy, $\Gamma^{e^+e^-}_\psi$ denotes the dilepton decay width of $\psi$ and $\Gamma_{\gamma \chi_{cJ}}$ represents the decay width for the $\psi \to \gamma \chi_{cJ}(2P)$ processes. Here, $m_\psi$ and $\Gamma_\psi$ refer to the mass and total width of the charmonium state $\psi$, respectively. For convenience, we will abbreviate these cross sections as $\sigma[\gamma \chi_{cJ}(2P)]$ in the following discussion.

   From Eq.~\eqref{sigma}, the relationship between the cross sections for the $\psi \to \gamma \chi_{cJ}(2P)$ processes is
	\begin{equation}
	\begin{split}
	    &\sigma[\gamma\chi_{c0}(2P)]:\sigma[\gamma\chi_{c1}(2P)]:\sigma[\gamma\chi_{c2}(2P)]\\
		&=\Gamma_{\gamma\chi_{c0}}:\Gamma_{\gamma\chi_{c1}}:\Gamma_{\gamma\chi_{c2}}.\label{012}
	\end{split}
	\end{equation}
	Once the ratios between $\Gamma_{\gamma \chi_{cJ}}$ are determined and $\sigma[\gamma \chi_{c1}(2P)]$ is extracted by fitting the data from the BESIII experiment~\cite{BESIII:2019qvy}, the cross sections $\sigma[\gamma \chi_{c0}(2P)]$ and $\sigma[\gamma \chi_{c2}(2P)]$ can be estimated. Therefore, we will focus on calculating the decay widths for the $\psi \to \gamma \chi_{cJ}(2P)$ processes.
	
    Given that the $\chi_{cJ}(2P)$ states exhibit a significant coupled-channel effect, we will use the hadronic loop mechanism to estimate the decay widths of $\psi \to \gamma \chi_{cJ}(2P)$. This effective approach has proven to be a powerful tool for modeling coupled-channel effects~\cite{Liu:2006dq,Liu:2009dr,Zhang:2009kr} and has been successfully applied to evaluate anomalous radiative transitions between $h_b(nP)$ and $\eta_b(mS)$~\cite{Chen:2013cpa}.

The diagrams depicting the $\psi \to \gamma \chi_{cJ}(2P)$ decay are shown in Fig.~\ref{diagram}. The triangle diagrams with loops of three charmed mesons represent the coupling between charmonium states and a pair of charmed mesons. Additionally, contact diagrams with two vertices are included to ensure gauge invariance, which will be discussed in detail later. The general form of the decay amplitudes can be expressed as
	\begin{align}
		&\mathcal{M}^{\rm{Tri}}=\int\frac{d^4q}{(2\pi)^4}\frac{\mathcal{V}_1\mathcal{V}_2\mathcal{V}_3}{D(q_1,m_{q_1})D(q_2,m_{q_2})D(q,m_{q})}\mathcal{F}^2(q^2),\\
		&\mathcal{M}^{\rm{Cont}}=\int\frac{d^4q_1}{(2\pi)^4}\frac{\mathcal{V}_1\mathcal{V}_2}{D(q_1,m_{q_1})D(q_2,m_{q_2})}\mathcal{F}^2_{\rm{cont}}(q_1^2),
	\end{align}
    respectively, where $D(p,m) = -i(p^2 - m^2)$ represents the propagator. The $\mathcal{V}_i$ denotes the interaction vertex, with the Lorentz indices omitted for simplicity. The form factors $\mathcal{F}(q^2)$ and $\mathcal{F}_{\rm{cont}}(q_1^2)$ are introduced to account for the off-shell effects of the exchanged charmed mesons and to regulate the ultraviolet (UV) divergences in the loop integrals. For $\mathcal{F}(q^2)$, we adopt the dipole form factor
    \begin{align}
    \mathcal{F}(q^2)=\left(\frac{m_E^2-\Lambda^2}{q^2-\Lambda^2}\right)^2,
    \end{align}
    where $m_E$ and $q$ represent the mass and four-momentum of the exchanged charmed mesons, respectively. The cut-off parameter is defined as $\Lambda = m_E + \alpha \Lambda_{\rm{QCD}}$, where $\Lambda_{\rm{QCD}} = 220$ MeV and $\alpha$ is a dimensionless free parameter, typically expected to be of order 1~\cite{Cheng:2004ru}. For the form factor $\mathcal{F}_{\rm{cont}}(q_1^2)$, it is introduced to ensure gauge invariance. This form factor cannot be arbitrary and is determined by the gauge invariance condition~\cite{Chen:2013cpa,Duan:2024zuo}.

    We use the effective Lagrangian approach to describe the diagrams in Fig.~\ref{diagram}. The Lagrangians depicting the interaction vertices between charmonium and a pair of $S$-wave charmed mesons are provided in~\cite{Casalbuoni:1996pg,Colangelo:2003sa,Xu:2016kbn,Chen:2013cpa,Duan:2024zuo}:
	\begin{eqnarray}
		\mathcal{L}_{\psi D^{(*)}D^{(*)}}&=&-ig_{\psi DD}\psi_{\mu}\left(\partial^{\mu} DD^{\dagger}-D\partial^{\mu}D^{\dagger}\right)\nonumber\\
		&&+g_{\psi D^*D}\varepsilon_{\mu\nu\alpha\beta}\partial^{\mu}\psi^{\nu}\left(D^{*\alpha}\overset{\leftrightarrow}{\partial^{\beta}}D^{\dagger}-D\overset{\leftrightarrow}{\partial^{\beta}}D^{*\alpha\dagger}\right)\nonumber\\
		&&+ig_{\psi D^{*}D^{*}}\psi^{\mu}\left(D^{*}_{\nu}\partial^{\nu}D^{*\dagger}_{\mu}-\partial^{\nu}D^{*}_{\mu}D^{*\dagger}_{\nu}-D^*_{\nu}\overset{\leftrightarrow}{\partial}_{\mu}D^{*\nu\dagger}\right), \nonumber\\
	\end{eqnarray}
	\begin{align}
		\mathcal{L}_{\chi_{cJ}D^{(*)}D^{(*)}}=&\;ig_{\chi_{c0}DD}\chi_{c0}DD^{\dagger}-ig_{\chi_{c0}D^*D^*}\chi_{c0}D^{*\mu}D^{*\dagger}_{\mu}\nonumber\\
		&-ig_{\chi_{c1}DD^*}\chi_{c1}^{\mu}\left(D^*_{\mu}D^{\dagger}-DD^{*\dagger}_{\mu}\right)\nonumber\\
		&+ig_{\chi_{c2}DD}\chi_{c2}^{\mu\nu}\partial_{\mu}D\partial_{\nu}D^{\dagger}+ig_{\chi_{c2}D^*D^*}\chi_{c2}^{\mu\nu}D^*_{\mu}D^{*\dagger}_{\nu}\nonumber\\
		&+g_{\chi_{c2}DD^*}\varepsilon_{\mu\nu\alpha\beta}\partial^{\alpha}\chi_{c2}^{\mu\rho}\left(\partial_{\rho}D^{*\nu}\partial^{\beta}D^{\dagger}-\partial^{\beta}D\partial_{\rho}D^{*\nu\dagger}\right).
	\end{align}
	The interaction Lagrangian describing the $D^{(*)}D^{(*)}\gamma$ vertices is given by~\cite{Chen:2013cpa,Duan:2024zuo}
	\begin{equation}
	\begin{split}
	    \mathcal{L}_{D^{(*)}D^{(*)}\gamma}=&\;i e A_{\mu}D^-\overset{\leftrightarrow}{\partial^{\mu}}D^{+}\\
		&+\frac{e\, g_{D^0\bar{D}^{*0}\gamma}}{4}\varepsilon^{\mu\nu\alpha\beta}F_{\mu\nu}\left(D^{*0}_{\alpha\beta}\bar{D}^0+\bar{D}^{*0}_{\alpha\beta}D^0\right)\\
		&+\frac{e\, g_{D^+D^{*-}\gamma}}{4}\varepsilon^{\mu\nu\alpha\beta}F_{\mu\nu}\left(D^{*+}_{\alpha\beta}D^-+D^{*-}_{\alpha\beta}D^+\right)\\
		&+i e A^{\mu}\bigg{(}g^{\alpha\beta}D^{*-}_{\alpha}\partial_{\beta}D^{*+}_{\mu}-g^{\alpha\beta}\partial_{\alpha}D^{*-}_{\mu}D^{*+}_{\beta} \\
		&+g^{\alpha\beta}D^*_{\alpha}\overset{\leftrightarrow}{\partial}_{\mu}D^{*+}_{\beta}\bigg{)}.
	\end{split}
	\end{equation}
    Then, the effective Lagrangians~\cite{Duan:2024zuo} describing the vertices where $\chi_{cJ}(2P)$ interacts with $D^{(*)}D^{(*)}\gamma$ are
	\begin{align}
		 &\mathcal{L}_{\chi_{cJ}D^{(*)}D^{(*)}\gamma}\nonumber\\
		 &=\; ie\, g_{\chi_{c0}DD\gamma}\chi_{c0}A_{\mu}D\overset{\leftrightarrow}{\partial^{\mu}}D^{\dagger}\nonumber\\
		&\quad+ie\, g_{\chi_{c0}D^*D^*\gamma}\chi_{c0}A^{\mu}\left(D^{*}_{\nu}\partial^{\nu}D^{*\dagger}_{\mu}-\partial^{\nu}D^{*}_{\mu}D^{*\dagger}_{\nu}+D^*_{\nu}\overset{\leftrightarrow}{\partial}_{\mu}D^{*\nu\dagger}\right)\nonumber\\
		&\quad-\frac{1}{\sqrt{2}}e\, f_{\chi_{c1}DD^*\gamma}A^{\mu}\chi_{c1}^{\alpha}\left( D^{*\dagger}_{\mu}\overset{\leftrightarrow}{\partial}_{\alpha}D-D^{\dagger}\overset{\leftrightarrow}{\partial}_{\alpha}D^*_{\mu} \right)\nonumber\\
		&\quad+ie\, g_{\chi_{c2}DD\gamma}\chi_{c2}^{\mu\nu}A_{\mu}D\overset{\leftrightarrow}{\partial}_{\nu}D^{\dagger}\nonumber\\
		&\quad+ie\, g_{\chi_{c2}DD^*\gamma}\varepsilon_{\mu\nu\alpha\beta}\partial^{\alpha}\chi_{c2}^{\mu\rho}A_{\rho}\left( D^{*\nu}\partial^{\beta}D^{\dagger}-\partial^{\beta}DD^{*\nu\dagger} \right)\nonumber\\
		&\quad+ie\, g_{\chi_{c2}D^*D^*\gamma}\partial^{\rho}\chi_{c2}^{\mu\nu}\partial_{\mu}A_{\nu}D^*_{\alpha}\overset{\leftrightarrow}{\partial}_{\rho}D^{*\alpha\dagger}.
	\end{align}
    Furthermore, the mass of the $\psi(4230)$ lies near the $D_1(2420)\bar{D}$ threshold. Crucially, the $D_1(2420)\bar{D}$ pair couples via $S$-wave interaction to form a $J^{PC} = 1^{--}$ configuration, implying substantial coupling between the $D_1(2420)\bar{D}$ channel and the $\psi(4230)$ state. Recent coupled-channel analyses~\cite{Man:2025zfu} have quantitatively validated this mechanism. Therefore, rigorously including contributions from the $D_1(2420)\bar{D}$ channel is imperative in studies of the $\psi \to \gamma \chi_{cJ}(2P)$ processes. The relevant interaction Lagrangians are given as follows:  
    \begin{equation}
	\begin{split}
	    \mathcal{L}_{\psi D_{1}D}=&\,ig_{\psi D_1D}\psi_{\mu}(D^{\mu\dagger}_1D-D^\dagger D_1^{\mu}),\\
		\mathcal{L}_{D_1D^{(*)}\gamma}=&\,ieg_{D_1D\gamma}A_\mu(D^\dagger D^\mu_1-D^{\mu\dagger}_1D)\nonumber\\
		&\,+ieg_{D_1D^*\gamma}\varepsilon_{\mu\nu\alpha\beta}F^{\mu\nu}(D^{*\alpha\dagger}D^{\beta}_1-D^{*\alpha}D_1^{\beta\dagger}),\\
		\mathcal{L}_{\chi_{c0}D_1D\gamma}=&\,ieg_{\chi_{c0}D_1D\gamma}\chi_{c0}A_{\mu}(D^\dagger D^\mu_1-D^{\mu\dagger}_1D),\\
		\mathcal{L}_{\chi_{c2}D_1D\gamma}=&\,ieg_{\chi_{c2}D_1D\gamma}\chi_{c2}^{\mu\nu}A_{\mu}(D^\dagger D_{1\nu}-D^\dagger_{1\nu}D).
	\end{split}
	\end{equation}
	Among these Lagrangians, $\overset{\leftrightarrow}{\partial}=\overset{\rightarrow}{\partial}-\overset{\leftarrow}{\partial}$, $D^*_{\alpha\beta}=\partial_{\alpha}D^*_{\beta}-\partial_{\beta}D^*_{\alpha}$, $F_{\mu\nu}=\partial_{\mu}A_{\nu}-\partial_{\nu}A_{\mu}$ and $e$ represents unit charge.
	
	For the $\psi \to \gamma \chi_{c1}(2P)$ decay, the corresponding amplitudes for the triangle diagrams can be written as
	\begin{align}
		\mathcal{M}^{(a)}_{\gamma\chi_{c1}}=&\;i^3\int\frac{d^4 q}{(2\pi)^4}\frac{1}{q^2_1-m^2_D}\frac{1}{q^2_2-m^2_D}\frac{-g^{\theta\tau}+q^{\theta}q^{\tau}/m^2_{D^*}}{q^2-m^2_{D^*}}\nonumber\\
		&\times\left[g_{\psi DD}\epsilon_{\psi}^{\mu}(q_{2\mu}-q_{1\mu})\right]\nonumber\\
		&\times\bigg[\frac{e\, g_{D^+D^{*-}\gamma}}{4}\varepsilon^{lnmt}(ip_{1l}g_{n\nu}-i p_{1n}g_{l\nu})\epsilon^{*\nu}_{\gamma}\nonumber\\
		&\times(i q_{m}g_{t\theta}- i q_{t}g_{m\theta})\bigg]\left[-i g_{\chi_{c1}DD^*}\epsilon^{*\kappa}_{\chi_{c1}}g_{\tau\kappa}\right]\mathcal{F}^2(q^2),
		\label{chi1a}
	\end{align}
	\begin{align}
		\mathcal{M}^{(b)}_{\gamma\chi_{c1}}=&\;i^3\int\frac{d^4 q}{(2\pi)^4}\frac{-g^{\alpha\beta}+q^{\alpha}_1q^{\beta}_1/m^2_{D^*}}{q^2_1-m^2_{D^*}}\frac{1}{q^2_2-m^2_D}\nonumber\\
		&\times\frac{-g^{\theta\tau}+q^{\theta}q^{\tau}/m^2_{D^*}}{q^2-m^2_{D^*}}\left[g_{\psi D^*D}\varepsilon_{\rho\mu\alpha\eta}p^{\rho}\epsilon_{\psi}^{\mu}(q^{\eta}_2-q^{\eta}_1)\right]\nonumber\\
		&\times\left[(-e\,\epsilon_{\gamma}^{*\nu})(g_{\beta\theta}(q_{\nu}+q_{1\nu})+g_{\theta\nu}q_{\beta}+g_{\beta\nu}q_{1\theta})\right]\nonumber\\
		&\times\left[-i g_{\chi_{c1}DD^*}\epsilon^{*\kappa}_{\chi_{c1}}g_{\tau\kappa}\right]\mathcal{F}^2(q^2),
	\end{align}
	\begin{align}
		\mathcal{M}^{(c)}_{\gamma\chi_{c1}}=&\;i^3\int\frac{d^4 q}{(2\pi)^4}\frac{1}{q^2_1-m^2_D}\frac{-g^{\delta\lambda}+q^{\delta}_2q^{\lambda}_2/m^2_{D^*}}{q^2_2-m^2_{D^*}}\frac{1}{q^2-m^2_D}\nonumber\\
		&\times\left[g_{\psi D^*D}\varepsilon_{\rho\mu\delta\eta}p^{\rho}\epsilon_{\psi}^{\mu}(q^{\eta}_1-q^{\eta}_2)\right]\nonumber\\
		&\times\left[- e\,\epsilon_{\gamma}^{*\nu}( q_{1\nu}+ q_{\nu})\right]\nonumber\\
		&\times\left[i g_{\chi_{c1}DD^*}\epsilon^{*\kappa}_{\chi_{c1}}g_{\lambda\kappa}\right]\mathcal{F}^2(q^2),
	\end{align}
	\begin{align}
		\mathcal{M}^{(d)}_{\gamma\chi_{c1}}=&\;i^3\int\frac{d^4 q}{(2\pi)^4}\frac{-g^{\alpha\beta}+q^{\alpha}_1q^{\beta}_1/m^2_{D^*}}{q^2_1-m^2_{D^*}}\frac{-g^{\delta\lambda}+q^{\delta}_2q^{\lambda}_2/m^2_{D^*}}{q^2_2-m^2_{D^*}}\nonumber\\
		&\times\frac{1}{q^2-m^2_D}\left[g_{\psi D^*D^*}\epsilon_{\psi}^{\mu}(q_{2\alpha}g_{\mu\delta}-q_{1\delta}g_{\mu\alpha}\right.\nonumber\\
		&\left.+(q_{1\mu}-q_{2\mu})g_{\alpha\delta})\right]\bigg[\frac{e\,  g_{D^+D^{*-}\gamma}}{4}\varepsilon^{lnmt}(ip_{1l}g_{n\nu}-i p_{1n}g_{l\nu})\nonumber\\
		&\times\epsilon^{*\nu}_{\gamma}(-i q_{1m}g_{t\beta}+ i q_{1t}g_{m\beta})\bigg]\left[i g_{\chi_{c1}DD^*}\epsilon^{*\kappa}_{\chi_{c1}}g_{\lambda\kappa}\right]\mathcal{F}^2(q^2).
		\label{chi1d}
	\end{align}
    \begin{align}
		\mathcal{M}^{(e)}_{\gamma\chi_{c1}}=&\;i^3\int\frac{d^4 q}{(2\pi)^4}\frac{-g^{\alpha\beta}+q^{\alpha}_1q^{\beta}_1/m^2_{D_1}}{q^2_1-m^2_{D_1}}\frac{1}{q^2_2-m^2_D}\frac{-g^{\theta\tau}+q^{\theta}q^{\tau}/m^2_{D^*}}{q^2-m^2_{D^*}}\nonumber\\
		&\times\left[i g_{\psi D_1D}\epsilon^{\mu}_\psi g_{\mu\alpha}\right]\nonumber\\
		&\times\left[-e\,g_{D_1^+D^{*-}\gamma}\varepsilon_{\rho\eta\theta\beta}(p_{1}^\rho g^{\eta\nu}-p_{1}^\eta g^{\rho\nu})\epsilon_{\gamma\nu}^{*}\right]\nonumber\\
		&\times\left[-i g_{\chi_{c1}DD^*}\epsilon^{*\kappa}_{\chi_{c1}}g_{\tau\kappa}\right]\mathcal{F}^2(q^2),
    \label{D1D}
    \end{align}
    To derive the amplitudes $\mathcal{M}^{(f)}_{\gamma\chi_{c1}}$ and $\mathcal{M}^{(g)}_{\gamma\chi_{c1}}$, we substitute the coupling constant $g_{D^+D^{*-}\gamma}$ with $g_{D^0\bar{D}^{*0}\gamma}$ in $\mathcal{M}^{(a)}_{\gamma\chi_{c1}}$ and $\mathcal{M}^{(d)}_{\gamma\chi_{c1}}$, respectively. Similarly, replacing $g_{D^+_1D^{*-}\gamma}$ with $g_{D^0_1\bar{D}^{*0}\gamma}$ in $\mathcal{M}^{(e)}_{\gamma\chi_{c1}}$ yields $\mathcal{M}^{(h)}_{\gamma\chi_{c1}}$.  

	In particular, for processes involving a photon, the amplitudes must satisfy gauge invariance. However, the decay amplitudes listed in Eqs. \eqref{chi1a}-\eqref{D1D} do not fully guarantee this. To address this issue, we introduce the contact diagrams shown in Fig.~\ref{diagram}~\cite{Chen:2013cpa,Duan:2024zuo}.

The amplitudes for the contact diagrams can be expressed as
    \begin{equation}
	\begin{split}
		\mathcal{M}^{(j)}_{\gamma\chi_{c1}}=&\;i^2\int\frac{d^4q_1}{(2\pi)^4}\frac{1}{q_1^2-m^2_D}\frac{-g^{\delta\lambda}+q^{\delta}_2q^{\lambda}_2/m^2_{D^*}}{q^2_2-m^2_{D^*}}\\
		&\times\left[-\frac{i}{\sqrt{2}}e\, f_{\chi_{c1}DD^*\gamma}\epsilon_{\gamma}^{*\nu}\epsilon_{\chi_{c1}}^{*\kappa}g_{\nu\lambda}(q_{2\kappa}-q_{1\kappa})\right]\\
		&\times\left[g_{\psi D^*D}\varepsilon_{\rho\mu\delta\eta}p^{\rho}\epsilon_{\psi}^{\mu}(q^{\eta}_1-q^{\eta}_2)\right]\mathcal{F}^2_{\rm{cont}}(q_1^2),
		\label{chi1e}
	\end{split}
	\end{equation}
	\begin{equation}
	\begin{split}
		\mathcal{M}^{(k)}_{\gamma\chi_{c1}}=&\;i^2\int\frac{d^4q_1}{(2\pi)^4}\frac{-g^{\alpha\beta}+q^{\alpha}_1q^{\beta}_1/m^2_{D^*}}{q^2_1-m^2_{D^*}}\frac{1}{q^2_2-m^2_D}\\
		&\times \left[\frac{i}{\sqrt{2}}e\, f_{\chi_{c1}DD^*\gamma}\epsilon_{\gamma}^{*\nu}\epsilon_{\chi_{c1}}^{*\kappa}g_{\nu\beta}(q_{2\kappa}-q_{1\kappa})\right]\\
		&\times\left[g_{\psi D^*D}\varepsilon_{\rho\mu\alpha\eta}p^{\rho}\epsilon_{\psi}^{\mu}(q^{\eta}_2-q^{\eta}_1)\right]\mathcal{F}^2_{\rm{cont}}(q_1^2).
		\label{chi1f}
	\end{split}
	\end{equation}
	
	Similarly, the amplitudes for $\psi \to \gamma \chi_{c0,2}(2P)$ can be easily written using the vertices listed in Appendix~\ref{A}.
The total amplitude for $\psi \to \gamma \chi_{c1}(2P)$ can then be expressed as
	\begin{align}
		\mathcal{M}^{\rm{Tri}}_{\gamma\chi_{c1}}=&\,2\sum_{i=a,b,c,d,e,f,g,h}\mathcal{M}^{(i)}_{\gamma\chi_{c1}},\\[6pt]
		\mathcal{M}^{\rm{Cont}}_{\gamma\chi_{c1}}=&\,2\sum_{i=j,k}\mathcal{M}^{(i)}_{\gamma\chi_{c1}},\\[6pt]
		\mathcal{M}^{\rm{Tot}}_{\gamma\chi_{c1}}=&\,\mathcal{M}^{\rm{Tri}}_{\gamma\chi_{c1}}+\mathcal{M}^{\rm{Cont}}_{\gamma\chi_{c1}},\label{tot}
	\end{align}	
	where the factor 2 comes from the sum over isospin doublet of charmed meson loops.

To ensure gauge invariance of the photon field, we need to determine the relationship between $\mathcal{M}^{\rm{Tri}}_{\gamma \chi_{c1}}$ and $\mathcal{M}^{\rm{Cont}}_{\gamma \chi_{c1}}$. Gauge invariance of the photon field is expressed through the Ward-Takahashi identity. For Feynman diagrams involving an external photon line, the total amplitude can always be written as $\mathcal{M}^{\rm{Tot}} = \epsilon_{\gamma}^{\mu} \mathcal{M}^{\rm{Tot}}_\mu$, where $\epsilon_{\gamma}^{\mu}$ is the polarization vector of the photon. The Ward-Takahashi identity implies that when the polarization vector of the photon is replaced by its four-momentum, the condition $p_1^{\mu} \mathcal{M}^{\rm{Tot}}_{\mu} = 0$ must hold, where $p_1$ is the photon four-momentum. To illustrate this principle, we will continue with the $\psi \to \gamma \chi_{c1}(2P)$ process as an example.

Examining Eq. \eqref{chi1a} through Eq. \eqref{chi1f}, we find that $\mathcal{M}^{(a)}_{\gamma \chi_{c1}}$, $\mathcal{M}^{(d)}_{\gamma \chi_{c1}}$, $\mathcal{M}^{(e)}_{\gamma \chi_{c1}}$, $\mathcal{M}^{(f)}_{\gamma \chi_{c1}}$, $\mathcal{M}^{(g)}_{\gamma \chi_{c1}}$ and $\mathcal{M}^{(h)}_{\gamma \chi_{c1}}$ satisfy the Ward-Takahashi identity directly. The remaining amplitudes can be expanded as a series of Lorentz structures with parameters after performing the loop integral:
    \begin{equation}
	\begin{split}
		\mathcal{M}^{(b)}_{\gamma\chi_{c1}}=&\;B_1\varepsilon_{\mu\nu\alpha\beta}p^{\mu}_1\epsilon_{\gamma}^{*\nu}\epsilon_{\chi_{c1}}^{*\alpha}\epsilon_{\psi}^{\beta}
		+B_2\varepsilon_{\mu\nu\alpha\beta}p^{\mu}_2\epsilon_{\gamma}^{*\nu}\epsilon_{\chi_{c1}}^{*\alpha}\epsilon_{\psi}^{\beta}\\
		&+B_3(p_1\cdot\epsilon_{\chi_{c1}}^*)\varepsilon_{\mu\nu\alpha\beta}p^{\mu}_1p^{\nu}_2\epsilon_{\gamma}^{*\alpha}\epsilon_{\psi}^{\beta}\\
		&+B_4(p_2\cdot\epsilon_{\gamma}^*)\varepsilon_{\mu\nu\alpha\beta}p^{\mu}_1p^{\nu}_2\epsilon_{\chi_{c1}}^{*\alpha}\epsilon_{\gamma}^{*\beta},\\
		\mathcal{M}^{(c)}_{\gamma\chi_{c1}}=&\;C_1\varepsilon_{\mu\nu\alpha\beta}p^{\mu}_1\epsilon_{\gamma}^{*\nu}\epsilon_{\chi_{c1}}^{*\alpha}\epsilon_{\psi}^{\beta}
		+C_2\varepsilon_{\mu\nu\alpha\beta}p^{\mu}_2\epsilon_{\gamma}^{*\nu}\epsilon_{\chi_{c1}}^{*\alpha}\epsilon_{\psi}^{\beta}\\
		&+C_3(p_2\cdot\epsilon_{\gamma}^*)\varepsilon_{\mu\nu\alpha\beta}p^{\mu}_1p^{\nu}_2\epsilon_{\chi_{c1}}^{*\alpha}\epsilon_{\gamma}^{*\beta},\\
		\mathcal{M}^{(g)}_{\gamma\chi_{c1}}=&\;G_1\varepsilon_{\mu\nu\alpha\beta}p^{\mu}_1\epsilon_{\gamma}^{*\nu}\epsilon_{\chi_{c1}}^{*\alpha}\epsilon_{\psi}^{\beta}
		+G_2\varepsilon_{\mu\nu\alpha\beta}p^{\mu}_2\epsilon_{\gamma}^{*\nu}\epsilon_{\chi_{c1}}^{*\alpha}\epsilon_{\psi}^{\beta},\\
		\mathcal{M}^{(h)}_{\gamma\chi_{c1}}=&\;H_1\varepsilon_{\mu\nu\alpha\beta}p^{\mu}_1\epsilon_{\gamma}^{*\nu}\epsilon_{\chi_{c1}}^{*\alpha}\epsilon_{\psi}^{\beta}+H_2\varepsilon_{\mu\nu\alpha\beta}p^{\mu}_2\epsilon_{\gamma}^{*\nu}\epsilon_{\chi_{c1}}^{*\alpha}\epsilon_{\psi}^{\beta},
	\end{split}	
	\end{equation}
	where $G_1 = G_2$ and $H_1 = H_2$. Clearly, the terms $\varepsilon_{\mu\nu\alpha\beta}p^{\mu}_1\epsilon_{\gamma}^{*\nu}\epsilon_{\chi_{c1}}^{*\alpha}\epsilon_{\psi}^{\beta}$ and $(p_2\cdot\epsilon_{\gamma}^*)\varepsilon_{\mu\nu\alpha\beta}p^{\mu}_1p^{\nu}_2\epsilon_{\chi_{c1}}^{*\alpha}\epsilon_{\gamma}^{*\beta}$ ensure gauge invariance, while the other terms do not. Using the identity $p_1^{\mu} \mathcal{M}^{\rm{Tot}}_{\mu} = 0$, we obtain
	\begin{align}
		G_2+H_2=-B_2-C_2+B_4(p_1\cdot p_2)+C_3(p_1\cdot p_2).
	\end{align}
	Thus, we can write $\mathcal{M}^{\rm{Cont}}_{\gamma \chi_{c1}}$ as
	\begin{equation}
	\begin{split}
		\mathcal{M}^{\rm{Cont}}_{\gamma\chi_{c1}}=&\;2[-B_2-C_2+B_4(p_1\cdot p_2)+C_3(p_1\cdot p_2)]\\
		&\times(\varepsilon_{\mu\nu\alpha\beta}p^{\mu}_1\epsilon_{\gamma}^{*\nu}\epsilon_{\chi_{c1}}^{*\alpha}\epsilon_{\psi}^{\beta}+\varepsilon_{\mu\nu\alpha\beta}p^{\mu}_2\epsilon_{\gamma}^{*\nu}\epsilon_{\chi_{c1}}^{*\alpha}\epsilon_{\psi}^{\beta}),
	\end{split}
	\end{equation}
	where the factor 2 also comes from the sum over isospin doublet of charmed meson loops. 
	From Eq.~\eqref{tot}, the total amplitude can be determined.

Finally, the decay width for $\psi \to \gamma \chi_{c1}(2P)$ can be evaluated as
    \begin{align}
		\Gamma_{\gamma\chi_{c1}}=\frac{1}{3}\frac{1}{8\pi}\frac{|\bm{p}_1|}{m^2_{\psi}}|\mathcal{M}^{\rm{Tot}}_{\gamma\chi_{c1}}|^2,
    \end{align}
	where $\bm{p}_1$ represents the center-of-mass momentum of the photon, and the factor of 1/3 accounts for the averaging over the polarization of the initial vector state $\psi$. The decay widths $\Gamma_{\gamma \chi_{c0}}$ and $\Gamma_{\gamma \chi_{c2}}$ can be calculated in a similar manner, and their ratios can then be determined.

\section{numerical results}\label{sec3}
    To calculate the ratios of $\Gamma_{\gamma\chi_{c0}}:\Gamma_{\gamma\chi_{c1}}:\Gamma_{\gamma\chi_{c2}}$, we first need to determine the various coupling constants in the Lagrangians. The coupling constants $g_{\psi D^{(*)}D^{(*)}}$ are determined using the $\psi(4230)$ partial widths reported in Ref.~\cite{Wang:2019mhs} as $g_{\psi DD} = 0.765$, $g_{\psi D^*D} = 0.054~\mathrm{GeV}^{-1}$, and $g_{\psi D^*D^*} = 1.320$. For the $g_{\psi D_1D}$, we use the $R_{\psi(4230)} = \Gamma^{e^+e^-}_{\psi(4230)} \cdot \mathcal{BR}(\psi(4230) \to \pi^+ D^{*-} D^0) = 2.70~\mathrm{eV}$~\cite{Wang:2023zxj} extracted from the measured cross section for $e^+e^- \to \pi^+ D^{*-} D^0$ at center-of-mass energies between 4.05 and 4.60 GeV~\cite{BESIII:2018iea} to estimate this coupling.
    Assuming the decay $\psi(4230) \to \pi^+ D^{*-} D^0$ proceed via the coupling of $\psi(4230)$ with virtual $\bar{D}_1(2420)^0$ and $D^0$, the amplitude is modeled as  
  \begin{align}
\mathcal{M}_{D_1D}=&\,g_{max}\epsilon_{\psi\mu}\frac{-g^{\mu\nu}+q^\mu q^\nu/m_{D_1}^2}{q^2-m^2_{D_1}+im_{D_1}\Gamma_{D_1}}\nonumber\\
&\times g_{D_1D^*\pi}(-3q_{4\nu}q_{4\lambda}+q^2_4 g_{\nu\lambda})\epsilon^{*\lambda}_{D^*},
    \end{align} 
    where $q$ is the four-momentum of the $D_1(2420)$, with $m_{D_1} = 2422.1~\mathrm{MeV}$, $\Gamma_{D_1} = 31.3~\mathrm{MeV}$, and $q_4$ is the four-momentum of the $\pi^+$.
    Here, the Lagrangians describing the $D_1(2420)D^*\pi$ is~\cite{Liu:2024ziu} 
    \begin{align}  
    \mathcal{L}_{D_1D\pi}=g_{D_1D^*\pi}(3D^\mu_1\partial_{\mu}\partial_\nu\pi D^{*\nu}-D^{\mu}_1\partial^{\nu}\partial_\nu\pi D^{*}_\mu).
    \end{align}
The coupling constant $g_{D_1D^*\pi}=12.67$ $\rm{GeV}^{-1}$ can be derived from $\Gamma(D_1(2420)\to D^*\pi)\approx \Gamma_{D_1}$.
    The three-body decay width $\Gamma_{\pi^+ D^{*-} D^0}$ is computed via  
\begin{align}  
  \Gamma_{\pi^+ D^{*-} D^0} = \int \frac{|\mathcal{M}_{D_1D}|^2}{(2\pi)^5 16 m_{\psi}^2} \, |\bm{q}_2| |\bm{q}_3^*| \, d\Omega_2 \, d\Omega_3^* \, dm_{D^*\pi},  
\end{align}  
where $m_{\psi} = 4222.1~\mathrm{MeV}$~\cite{ParticleDataGroup:2022pth}, $\bm{q}_2$ and $\Omega_2$ denote the momentum and solid angle of $D^0$ in the $\psi$ rest frame, and $\bm{q}_3^*$, $\Omega_3^*$ represent the momentum and solid angle of $D^{*-}$ in the $D^{*-}\pi^+$ center-of-mass frame. Using $\Gamma^{e^+e^-}_{\psi(4230)} = 0.290~\mathrm{keV}$~\cite{Wang:2019mhs} and $\Gamma_{\psi(4230)} = 49~\mathrm{MeV}$~\cite{ParticleDataGroup:2022pth}, we find $g_{\mathrm{max}} = 10.19~\mathrm{GeV}^{-1/2}$. However, this overestimates the true coupling $g_{\psi D_1D}$, as additional intermediate states contribute to $\pi^+ D^{*-} D^0$, including $D^{*-}D^{*}_0(2300)^+$, $D^0\bar{D}_1(2430)$, and $D^0\bar{D}^{*}_2(2460)^0$. Coupled-channel studies~\cite{Man:2025zfu} confirm significant couplings between $\psi(4230)$ and these intermediate states. To account for their collective contributions, we adopt $g_{\psi D_1D} = g_{\mathrm{max}}/4 = 2.55~\mathrm{GeV}^{-1/2}$ in subsequent calculations.  

Specifically, the couplings of the photon with $S$-wave charmed mesons are given by $g_{D^0 \bar{D}^{*0} \gamma} = 2.0$ $\rm{GeV}^{-1}$ and $g_{D^+ D^{*-} \gamma} = -0.5$ $\rm{GeV}^{-1}$~\cite{Chen:2010re}. $g_{D^0_1\bar{D}^0\gamma}=1.446$ $\rm{GeV}^{-1/2}$, $g_{D^+_1 D^-\gamma}=0.466$ $\rm{GeV}^{-1/2}$, $g_{D^0_1\bar{D}^{*0}\gamma}=0.610$ and $g_{D^+_1\bar{D}^{*-}\gamma}=0.200$ are determined by the corresponding widths from Ref.~\cite{Godfrey:2015dva}.

Finally, the coupling constants connecting the $P$-wave charmonium multiplet to charmed mesons are expressed in terms of a universal coupling constant \( g_p \)~\cite{Li:2021jjt}:
    \begin{align}
       \frac{g_{\chi_{c0}DD}}{\sqrt{3}m_D}&=\frac{\sqrt{3}g_{\chi_{c0}D^*D^*}}{m_{D^*}}=2\sqrt{m_{\chi_{c0}}}g_p,\nonumber\\
		g_{\chi_{c1}DD^*}&=2\sqrt{2}\sqrt{m_Dm_{D^*}m_{\chi_{c1}}}g_p,\nonumber\\
		g_{\chi_{c2}DD}m_D&=g_{\chi_{c2}DD^*}\sqrt{m_Dm_{D^*}}m_{\chi_{c2}}=\frac{g_{\chi_{c2}D^*D^*}}{4m_{D^*}}=\sqrt{m_{\chi_{c2}}}g_p.
	\end{align} 
 The masses of all participating states in these relations are provided in Table~\ref{mass}, with values taken from the PDG~\cite{ParticleDataGroup:2022pth}.
        \renewcommand{\arraystretch}{1.20}
	\begin{table}
	\caption{The masses of states used in the calculation.}\label{mass}
	\begin{ruledtabular}
	\begin{tabular}{c|d|c|d}
		\textrm{States} & \multicolumn{1}{c|}{\textrm{Mass (MeV)}} & \textrm{States} & \multicolumn{1}{c}{\textrm{Mass (MeV)}} \\\hline
		$\psi(4230)$ & 4222.1 & $X(3915)\equiv\chi_{c0}(2P)$ & 3922.1 \\
		$X(3872)\equiv\chi_{c1}(2P)$ & 3871.64 & $Z(3930)\equiv\chi_{c2}(2P)$ & 3922.5\\
		$D^{\pm}$ & 1869.66 & $D^0$ & 1864.84\\
		$D^{*\pm}$ & 2010.26 & $D^{*0}$ & 2006.85\\
            $D_1$ & 2422.1 & $\pi^+$ & 139.57\\
	\end{tabular}
	\end{ruledtabular}
	\end{table}
   
\begin{figure}[htbp]
		\centering
		\includegraphics[width=8.7cm]{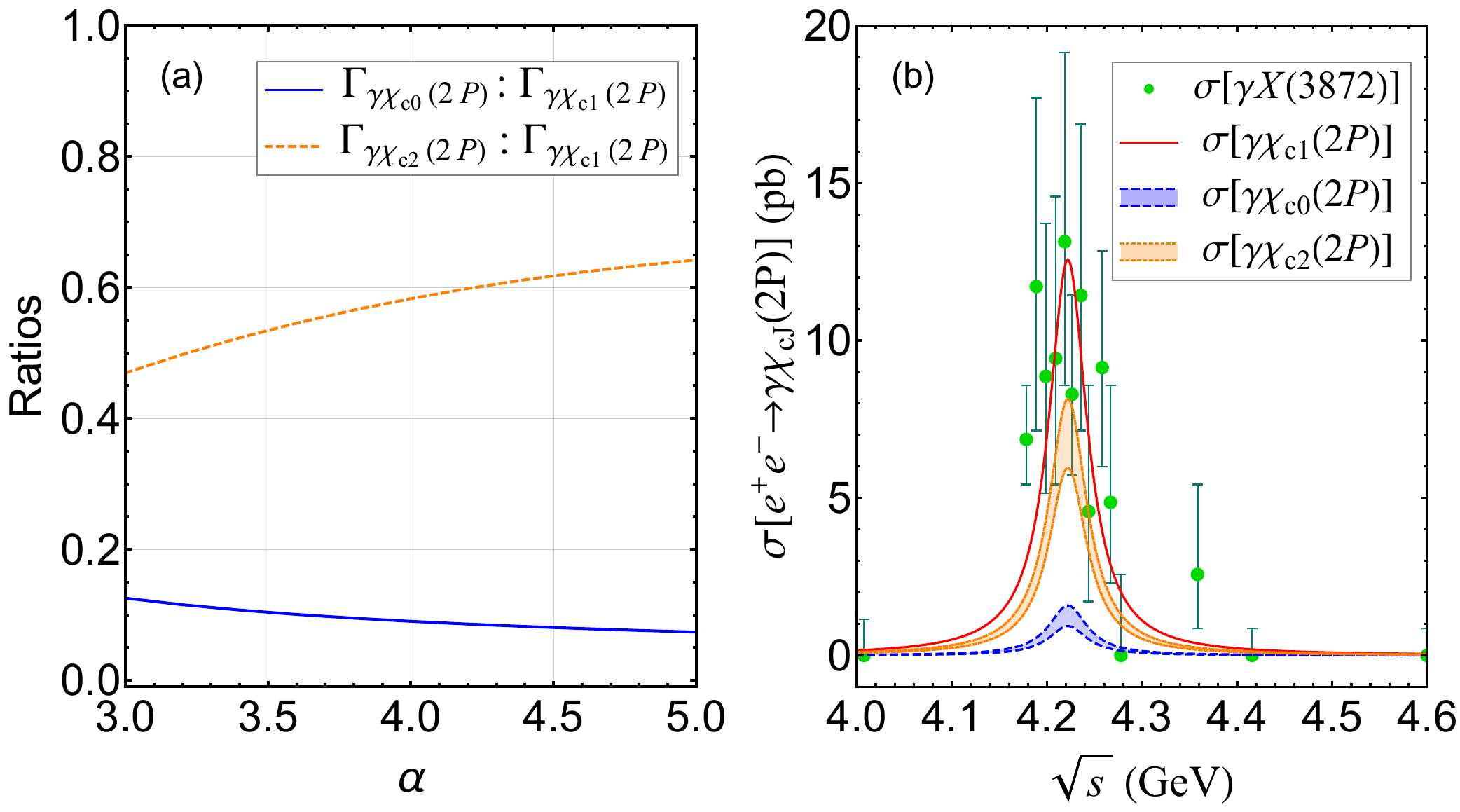}
		\caption{(a) The ratios of the partial widths for the three decay channels $\psi \to \gamma \chi_{cJ}(2P)$ as a function of $\alpha$.
(b) Treating $X(3872) \equiv \chi_{c1}(3872)$, the predicted cross sections for $e^+e^- \to \gamma \chi_{c0}(2P)$ and $e^+e^- \to \gamma \chi_{c2}(2P)$. The data points are obtained by dividing the original BESIII experiment data~\cite{BESIII:2019qvy} by the branching fraction of $X(3872) \to \pi^+ \pi^- J/\psi$. }
		\label{rs}
    \end{figure}

    Although the coupling constant $g_p$ are unknown, their relative ratios are well-defined. As a result, the cut-off parameter $\alpha$ becomes the only free parameter in our calculations. It is important to examine the impact of $\alpha$ on the ratios of $\Gamma_{\gamma \chi_{cJ}}$. Note that the dipole form factor used in the amplitudes resembles propagators with mass $\Lambda$, which introduces an additional branch cut in the loop integral when two propagators can be on-shell simultaneously, i.e., when $\Lambda + m_{D^{(*)}} = m_{\chi_{cJ}(2P)}$. Therefore, we choose $\alpha$ to lie in the range of 3 to 5 to avoid this condition, and the resulting ratios are shown in Fig.~\ref{rs}(a).

    It is observed that the ratios vary smoothly with $\alpha$. All partial widths of the three decay channels $\psi \to \gamma \chi_{cJ}(2P)$ remain within the same order of magnitude, with $\Gamma_{\gamma\chi_{c1}}$ is slightly larger than $\Gamma_{\gamma\chi_{c0}}$ and $\Gamma_{\gamma\chi_{c2}}$. The obtained ratios of the different decay widths are
    \begin{equation}
	\begin{split}
		\Gamma_{\gamma\chi_{c0}}:\Gamma_{\gamma\chi_{c1}}&=0.07\sim0.13 \,,\\
		\Gamma_{\gamma\chi_{c2}}:\Gamma_{\gamma\chi_{c1}}&=0.47\sim0.64 \,.
	\end{split}
    \end{equation}
    For the comparative analysis, we evaluate the ratios by systematically varying the coupling constant \( g_{\psi D_1D} \) as \( g_{\psi D_1D} = g_{\text{max}} \) and \( g_{\psi D_1D} = g_{\text{max}}/2 \), while setting the cut-off parameter \( \alpha \) to typical values of 3, 4, and 5. The results obtained under these parameter configurations are summarized in Table~\ref{gmax}. Notably, the ratios exhibit a significant dependence on the value of the coupling constant \( g_{\psi D_1D} \). However, a precise determination of \( g_{\psi D_1D} \) necessitates additional experimental measurements.
    
    \begin{table}[htbp]
    \caption{The ratios of three partial widths with different coupling constant $g_{\psi D_1D}$ and cut-off parameter $\alpha$.}\label{gmax}
        \setlength{\tabcolsep}{0.6cm}{
        \begin{tabular}{c|c|c|c}
        \hline\hline
                \multicolumn{4}{c}{$g_{\psi D_1D}=g_{max}$} \\ \hline
			$\alpha$ & 3 & 4 & 5 \\ \hline
			$\Gamma_{\gamma\chi_{c0}}:\Gamma_{\gamma\chi_{c1}}$ & 0.018 & 0.011 & 0.008 \\ \hline
                $\Gamma_{\gamma\chi_{c2}}:\Gamma_{\gamma\chi_{c1}}$ & 0.029 & 0.037 & 0.042 \\ \hline
                \multicolumn{4}{c}{$g_{\psi D_1D}=g_{max}/2$} \\ \hline
                $\alpha$ & 3 & 4 & 5 \\ \hline
                $\Gamma_{\gamma\chi_{c0}}:\Gamma_{\gamma\chi_{c1}}$ & 0.040 & 0.027 & 0.022 \\ \hline
                $\Gamma_{\gamma\chi_{c2}}:\Gamma_{\gamma\chi_{c1}}$ & 0.118 & 0.149 & 0.168 \\
        \hline\hline
	\end{tabular}
        }
    \end{table}

    To estimate $\sigma[\gamma \chi_{c0}(2P)]$ and $\sigma[\gamma \chi_{c2}(2P)]$, we first need to extract $\sigma[\gamma \chi_{c1}(2P)]$ from experimental data. It is important to note that the BESIII Collaboration has measured the cross section $\sigma(e^+e^- \to \gamma X(3872) \to \gamma \pi^+ \pi^- J/\psi)$~\cite{BESIII:2019qvy}. This cross section can be expressed as
    \begin{equation}
	\begin{split}
		&\sigma(e^+e^-\rightarrow\gamma X(3872)\rightarrow\gamma\pi^+\pi^-J/\psi)\\
		&=\sigma[\gamma X(3872)]\times\mathcal{BR}(X(3872)\rightarrow\pi^+\pi^-J/\psi).
        \label{sigma2}
	\end{split}
	\end{equation}
	By using Eq.~\eqref{sigma} and Eq.~\eqref{sigma2}, we attempt to provide a simple description of the BESIII data. Ignoring background contributions, we find that the cross-section data for $\sigma(\gamma X(3872) \to \gamma \pi^+ \pi^- J/\psi)$ can be well explained by the contribution of the $\psi(4230)$. The mass and width of $\psi(4230)$ are fixed at 4222.1 MeV and 49 MeV, respectively, based on values from the PDG~\cite{ParticleDataGroup:2022pth}. Although cross-section data for the process $e^+e^- \to \gamma X(3872) \to \gamma \omega J/\psi$ have also been provided by the BESIII experiment~\cite{BESIII:2019qvy}, fitting these data using a similar approach has been less satisfactory, possibly due to measurement errors.

Taking the branching ratio $\mathcal{BR}(X(3872) \to \pi^+ \pi^- J/\psi) = 3.5\%$ from the PDG~\cite{ParticleDataGroup:2022pth}, we can extract the cross section $\sigma[\gamma \chi_{c1}(2P)]$. Using the relation from Eq.~\eqref{012}, we estimate the cross sections $\sigma[\gamma \chi_{c0}(2P)]$ and $\sigma[\gamma \chi_{c2}(2P)]$, which are shown in Fig.~\ref{rs} (b). The cross sections $\sigma[\gamma \chi_{c1}(2P)]$ and $\sigma[\gamma \chi_{c2}(2P)]$ are sufficiently large to be measured experimentally, with $\sigma[\gamma \chi_{c2}(2P)]<\sigma[\gamma \chi_{c1}(2P)]$, while $\sigma[\gamma \chi_{c0}(2P)]$ is relatively hard to be measured. We expect that future experiments will measure these cross sections precisely, such as those from BESIII and Belle II.

With above results, we now focus on the $D\bar{D}$ decay channel of $\chi_{c0}(2P)$ and $\chi_{c2}(2P)$. For $\chi_{c0}(2P)$, $D\bar{D}$ is the only allowed open-charm decay mode, while for $\chi_{c2}(2P)$, $D\bar{D}$ constitutes approximately 60\% of the total decay rate~\cite{Liu:2009fe,Chen:2013yxa,Chen:2012wy,Duan:2020tsx,Duan:2021bna}. This makes the $e^+e^- \to \gamma D\bar{D}$ process an excellent avenue for the simultaneous observation of $\chi_{c0}(2P)$ and $\chi_{c2}(2P)$.

\begin{figure}[htbp]
		\centering
		\includegraphics[width=8.5cm]{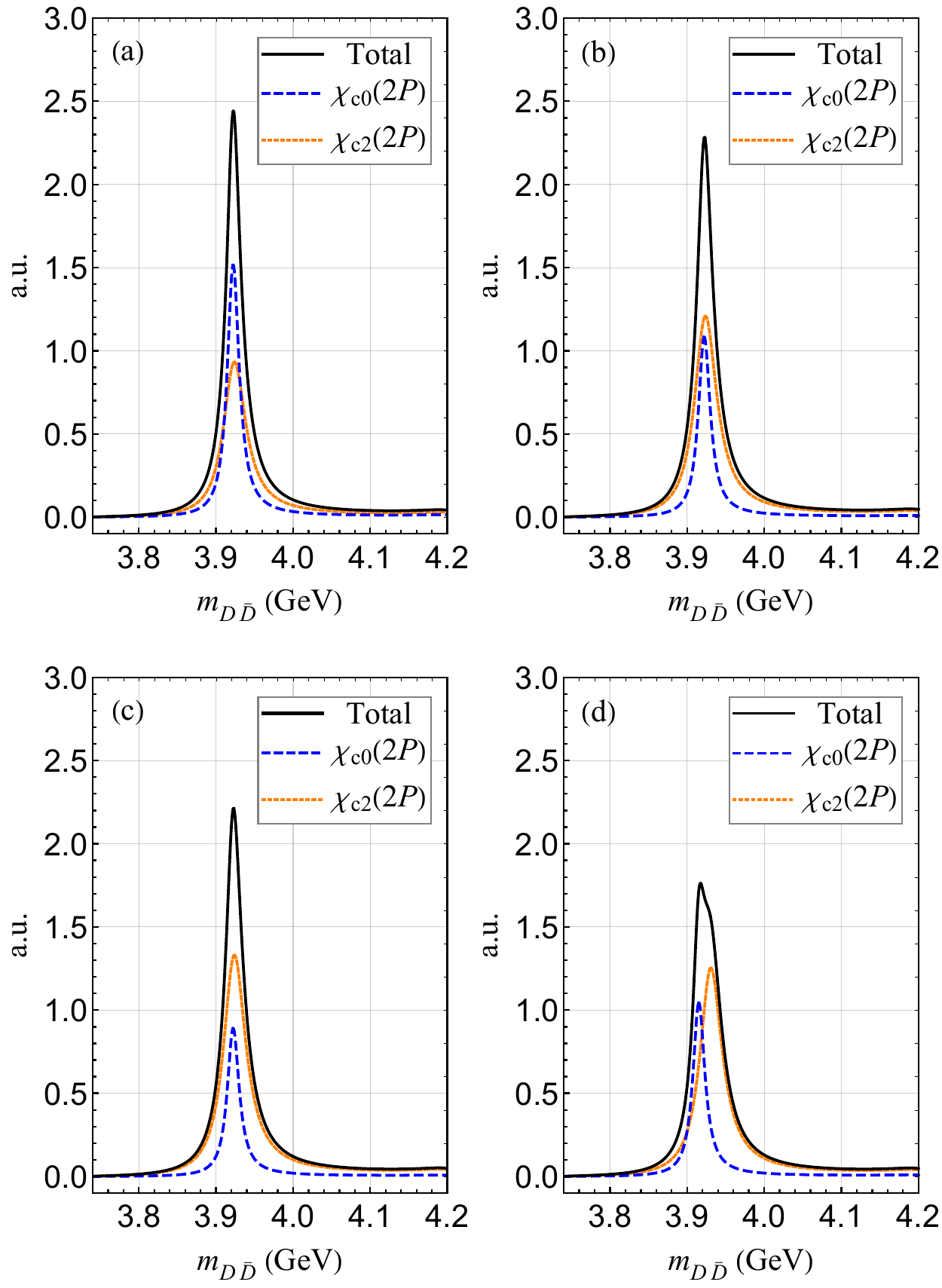}
		\caption{(a)-(c) The predicted $D\bar{D}$ invariant mass spectrum for $e^+e^- \to \gamma D\bar{D}$ with the cut-off parameter $\alpha$ set to 3, 4, and 5, respectively. (d) Revisiting the $D\bar{D}$ invariant mass spectrum of $e^+e^- \to \gamma D\bar{D}$ for $\alpha = 4$ with a slightly increased mass gap.}
		\label{4DD}
    \end{figure}

Assuming 100\% branching ratio for $\chi_{c0}(2P) \to D\bar{D}$ and 60\% for $\chi_{c2}(2P) \to D\bar{D}$, we depict the $D\bar{D}$ invariant mass spectrum for $e^+e^- \to \gamma D\bar{D}$ in Fig.~\ref{4DD} (a)-(c), with the cut-off parameter $\alpha$ set to typical values of 3, 4, and 5, respectively. The masses of $\chi_{c0}(2P)$ and $\chi_{c2}(2P)$ are taken as 3922.1 MeV and 3922.5 MeV, respectively. It is observed that the total amplitudes exhibit minimal variation across different values of $\alpha$. Although both states show significant amplitudes in the $D\bar{D}$ invariant mass spectrum, the small mass gap between $\chi_{c0}(2P)$ and $\chi_{c2}(2P)$ presents a challenge in distinguishing them.

In a preliminary investigation, we explore the effect of increasing the mass gap between $\chi_{c0}(2P)$ and $\chi_{c2}(2P)$ by adjusting their masses to 3915 MeV~\cite{Belle:2009and} and 3930 MeV~\cite{Belle:2005rte}, respectively. With $\alpha$ set to 4, we re-examine the $D\bar{D}$ invariant mass spectrum, as shown in Fig.~\ref{4DD} (d). Despite this adjustment, distinguishing between $\chi_{c0}(2P)$ and $\chi_{c2}(2P)$ remains a challenging task. We suggest that future experiments should focus on this process for further insights, such as those conducted by BESIII and Belle II.

    To aid experimental efforts in distinguishing between $\chi_{c0}(2P)$ and $\chi_{c2}(2P)$, we analyze the angular distributions of the $\psi \to \gamma \chi_{c0,2}(2P)$ processes. In the cases of $\psi \to \gamma \chi_{c0,2}(2P)$, if the produced charmonium is unpolarized, the angular distributions would be uniform. However, the $J^{PC} = 1^{--}$ charmonium produced directly from $e^+e^-$ annihilation is polarized, the corresponding spin density matrix is given by $\rho_{\psi} = \frac{1}{2} \sum_{m=\pm 1} |1,m \rangle \langle 1,m|$. As a result, the angular distributions of the $\psi \to \gamma \chi_{c0,2}(2P)$ processes can be expressed as
    \begin{align}
      \frac{dN}{Nd\cos\theta}\propto1+\beta_{0,2}\cos^2\theta,   
    \end{align}
    where $\theta$ denotes the polar angle of the photon in the center-of-mass frame of the $e^+e^-$ system. The coefficients $\beta_{0,2}$ are associated with $\chi_{c0}$ and $\chi_{c2}$, respectively, and are typically determined by the decay amplitudes. For the $\psi \to \gamma \chi_{c0}(2P)$ process, the angular distribution is given by
	\begin{equation}
	\begin{split}
	\frac{d\Gamma_{\gamma\chi_{c0}}}{d\Omega}&\propto\sum_{m,m'}\sum_{\mu=\pm 1} \rho_{m,m'}D^{j*}_{m \mu}(\phi,\theta,0)D^j_{m'\mu}(\phi,\theta,0)T^{j*}_\mu T^j_\mu \\
	&=\frac{1}{2}|T_1|^2(1+\cos^2\theta),
	\end{split}
	\end{equation}
    where $T_1 = T_{-1}$, $T^j_\mu$ denotes the helicity amplitude, and $D^j_{m^{(')}\mu}(\phi,\theta,0)$ represents the Wigner-D matrix. The angular distributions parameter $\beta_0$ for the $\psi \rightarrow \gamma \chi_{c0}(2P)$ process is fixed at 1, independent of the model. For the $\psi \rightarrow \gamma \chi_{c2}(2P)$ process, considering the spin density matrix $\rho_{\psi}$, the angular distribution parameter $\beta_2$ can be calculated directly from the differential partial width. This is given by the following formula:
    \begin{align}
		\frac{d\Gamma_{\gamma\chi_{c2}}}{d\Omega}=\frac{1}{32\pi^2}\frac{|\bm{p}_1|}{m^2_{\psi}}|\mathcal{M}^{\rm{Tot}}_{\gamma\chi_{c2}}|^2.
		\label{theta}
    \end{align}
    With the cut-off parameter $\alpha$ set to typical values of 3, 4, and 5, the obtained angular distributions parameters $\beta_2$ are listed in Table~\ref{a}. The parameter $\beta_2$ shows minimal dependence on variations in the cut-off parameter $\alpha$, and the ratio between $\beta_0$ and $\beta_2$ is approximately 9. We hope that these results will be useful in experimentally distinguishing between the $\chi_{c0}(2P)$ and $\chi_{c2}(2P)$ states.
    
    \begin{table}[htbp]
    \caption{$\beta_2$ for different values of $\alpha$.}\label{a}
        \setlength{\tabcolsep}{0.7cm}{
        \begin{tabular}{c|c|c|c}
        \toprule\toprule
			$\alpha$ & 3 & 4 & 5 \\ \hline
			$\beta_2$ & 0.109 & 0.110 & 0.112 \\
		\bottomrule\bottomrule
	\end{tabular}
        }
    \end{table}

\section{Discussions and conclusions}

	The study of the $\chi_{cJ}(2P)$ states is essential for deepening our understanding of charmonium properties, as highlighted by the historical development of this triplet. In particular, investigating the production modes of $\chi_{cJ}(2P)$ is crucial. Notably, the BESIII Collaboration's study of the $e^+e^- \rightarrow \gamma X(3872)$ process~\cite{BESIII:2019qvy} has motivated us to propose that the final state $\gamma X(3872)$ could originate from the radiative decay of the intermediate charmonium state $\psi(4230)$. In this context, the $e^+e^- \rightarrow \gamma \chi_{cJ}(2P)$ processes emerge as promising production channels for the $\chi_{cJ}(2P)$ states.

Considering the significant coupled-channel effects in these charmonium states, we have employed the hadronic loop mechanism~\cite{Liu:2006dq,Liu:2009dr,Zhang:2009kr,Chen:2013cpa,Casalbuoni:1996pg,Colangelo:2003sa,Xu:2016kbn,Duan:2024zuo,Duan:2024zuo,Li:2021jjt,Chen:2010re,Cheng:2004ru} to calculate the decay width ratios for the $\psi \rightarrow \gamma \chi_{cJ}(2P)$ processes. By treating $X(3872)$ as the charmonium state $\chi_{c1}(2P)$ and scaling the cross-section data from the BESIII experiment~\cite{BESIII:2019qvy}, we have determined the cross sections for $e^+e^- \rightarrow \gamma \chi_{c0}(2P)$ and $e^+e^- \rightarrow \gamma \chi_{c2}(2P)$. The calculated cross sections are found to be of similar magnitude and fall within the measurable range of future experiments, such as those at BESIII and Belle II.

To further investigate the properties of $\chi_{c0}(2P)$ and $\chi_{c2}(2P)$, we focused on their dominant decay channel, $D\bar{D}$~\cite{Liu:2009fe,Chen:2012wy}. We analyzed the $e^+e^- \rightarrow \gamma \chi_{c0,2}(2P) \rightarrow \gamma D\bar{D}$ processes and predicted the corresponding cross sections, $D\bar{D}$ invariant mass spectra, and angular distributions for the $\psi \rightarrow \gamma \chi_{c0,2}(2P)$ decays. These predictions are intended to guide future experimental efforts.

In conclusion, our study demonstrates that the $e^+e^- \rightarrow \gamma D\bar{D}$ process offers an excellent opportunity to observe the $\chi_{c0}(2P)$ and $\chi_{c2}(2P)$ states. We strongly recommend that experimental collaborations, such as BESIII and Belle II, prioritize this process for future investigations.

\begin{acknowledgments}
This work is supported by the National Natural Science Foundation of China under Grant Nos. 12335001, 12247101 and 12447124, the ‘111 Center’ under Grant No. B20063, the Natural Science Foundation of Gansu Province (No. 22JR5RA389), the fundamental Research Funds for the Central Universities, and the project for top-notch innovative talents of Gansu province.

\end{acknowledgments}

\appendix

\section{Feynman rules}\label{A}
In this appendix, we collect the Feynman rules for each interaction vertex as follows,
\begin{align}
	&\langle D(q_1)\bar{D}(q_2)|\psi(p)\rangle=-g_{\psi DD}\epsilon_{\psi}^{\mu}(q_{1\mu}-q_{2\mu}), 
\end{align}
\begin{align}
	&\langle D(q_1)\bar{D}^*(q2)|\psi(p)\rangle=g_{\psi DD^*}\varepsilon_{\mu\nu\alpha\beta}p^{\mu}\epsilon_{\psi}^{\nu}(q^{\alpha}_2-q^{\alpha}_1)\epsilon_{\bar{D}^*}^{*\beta},
\end{align}
\begin{align}
	&\langle D^*(q_1)\bar{D}(q_2)|\psi(p)\rangle=g_{\psi DD^*}\varepsilon_{\mu\nu\alpha\beta}p^{\mu}\epsilon_{\psi}^{\nu}(q^{\alpha}_1-q^{\alpha}_2)\epsilon_{\bar{D}^*}^{*\beta},
\end{align}
\begin{align}
	\langle D^*(q_1)\bar{D}^*(q_2)|\psi(p)\rangle=&\;g_{\psi D^*D^*}\epsilon_{\psi}^{\mu}(g_{\mu\beta}q_{2\alpha}-g_{\mu\alpha}q_{1\beta}\nonumber\\
	&+g_{\alpha\beta}(q_{1\mu}-q_{2\mu}))\epsilon_{D^*}^{*\alpha}\epsilon_{\bar{D}^*}^{*\beta},
\end{align}
\begin{align}
	\langle\chi_{c0}(p_2)|D(q)\bar{D}(q_2)\rangle=ig_{\chi_{c0}DD},
\end{align}
\begin{align}
	\langle\chi_{c0}(p_2)|D^*(q)\bar{D}^*(q_2)\rangle=-ig_{\chi_{c0}D^*D^*}g_{\alpha\beta}\epsilon_{D^*}^{\alpha}\epsilon_{\bar{D}^*}^{\beta},
\end{align}
\begin{align}
	\langle\chi_{c1}(p_2)|D^*(q)\bar{D}(q_2)\rangle=-ig_{\chi_{c1}DD^*}g_{\mu\alpha}\epsilon_{\chi_{c1}}^{*\mu}\epsilon_{D^*}^{\alpha},
\end{align}
\begin{align}
	\langle\chi_{c1}(p_2)|D(q)\bar{D}^*(q_2)\rangle=ig_{\chi_{c1}DD^*}g_{\mu\alpha}\epsilon_{\chi_{c1}}^{*\mu}\epsilon_{\bar{D}^*}^{\alpha},
\end{align}
\begin{align}
	\langle\chi_{c2}(p_2)|D(q)\bar{D}(q_2)\rangle=-ig_{\chi_{c2}DD}\epsilon_{\chi_{c2}}^{*\mu\nu}q_{\mu}q_{2\nu},
\end{align}
\begin{align}
	\langle\chi_{c2}(p_2)|D^*(q)\bar{D}(q_2)\rangle=ig_{\chi_{c2}DD^*}\varepsilon_{\mu\rho\alpha\beta}\epsilon_{\chi_{c2}}^{*\mu\nu}p^{\rho}_2q_{\nu}q^{\beta}_2\epsilon_{D^*}^{\alpha},
\end{align}
\begin{align}
	\langle\chi_{c2}(p_2)|D(q)\bar{D}^*(q_2)\rangle=-ig_{\chi_{c2}DD^*}\varepsilon_{\mu\rho\alpha\beta}\epsilon_{\chi_{c2}}^{*\mu\nu}p^{\rho}_2q_{2\nu}q^{\beta}\epsilon_{\bar{D}^*}^{\alpha},
\end{align}
\begin{align}
	\langle\chi_{c2}(p_2)|D^*(q)\bar{D}^*(q_2)\rangle=ig_{\chi_{c2}D^*D^*}g_{\mu\alpha}g_{\nu\beta}\epsilon_{\chi_{c2}}^{*\mu\nu}\epsilon_{D^*}^{\alpha}\epsilon_{\bar{D}^*}^{\beta},
\end{align}
\begin{align}
	\langle\gamma(p_1)D(q)|D(q_1)\rangle=-e\, \epsilon_{\gamma}^{*\mu}(q_{1\mu}+q_{\mu}),
\end{align}
\begin{align}
	\langle\gamma(p_1)D^*(q)|D(q_1)\rangle=&-\frac{e\, g_{DD^*\gamma}}{4}\varepsilon^{\lambda\nu\rho\beta}(p_{1\lambda}g_{\nu\mu}-p_{1\nu}g_{\lambda\mu})\epsilon^{*\mu}_{\gamma}\nonumber\\
	&\times(q_{\rho}g_{\beta\alpha}- q_{\beta}g_{\rho\alpha})\epsilon_{D^*}^{*\alpha},
\end{align}
\\
\begin{align}
	\langle\gamma(p_1)D(q)|D^*(q_1)\rangle=&-\frac{e\,  g_{DD^*\gamma}}{4}\varepsilon^{\lambda\nu\rho\beta}(p_{1\lambda}g_{\nu\mu}-p_{1\nu}g_{\lambda\mu})\epsilon^{*\mu}_{\gamma}\nonumber\\
	&\times(- q_{1\rho}g_{\beta\alpha}+ q_{1\beta}g_{\rho\alpha})\epsilon_{D^*}^{\alpha},
\end{align}
\begin{align}
	\langle\gamma(p_1)D^*(q)|D^*(q_1)\rangle=&\;(-e\,\epsilon_{\gamma}^{*\mu})(g_{\beta\alpha}(q_{\mu}+q_{1\mu})+g_{\alpha\mu}q_{\beta}\nonumber\\
	&+g_{\beta\mu}q_{1\alpha})\epsilon_{D^*}^{*\alpha}\epsilon_{D^*}^{\beta},
\end{align}
\begin{align}
	\langle\gamma(p_1)\chi_{c0}(p_2)|D(q_1)\bar{D}(q_2)\rangle=e\, g_{\chi_{c0}DD\gamma}\epsilon_{\gamma}^{\mu}(q_{2\mu}-q_{1\mu}),
\end{align}
\begin{align}
	\langle\gamma(p_1)\chi_{c0}(p_2)|D^*(q_1)\bar{D}^*(q_2)\rangle=&\;e\, g_{\chi_{c0}D^*D^*\gamma}\epsilon_{\gamma}^{\mu}(g_{\alpha\beta}(q_{2\mu}-q_{1\mu})\nonumber\\
	&+g_{\beta\mu}q_{2\alpha}-g_{\alpha\mu}q_{1\beta})\epsilon_{D^*}^{\alpha}\epsilon_{\bar{D}^*}^{\beta},
\end{align}
\begin{align}
	\langle\gamma(p_1)\chi_{c1}(p_2)|D(q_1)\bar{D}^*(q_2)\rangle=&-\frac{i}{\sqrt{2}}e\, f_{\chi_{c1}DD^*\gamma}\epsilon_{\gamma}^{*\mu}\epsilon_{\chi_{c1}}^{*\nu}g_{\mu\alpha}\nonumber\\
	&\times(q_{2\nu}-q_{1\nu})\epsilon_{\bar{D}^*}^{\alpha},
\end{align}

\begin{align}
	\langle\gamma(p_1)\chi_{c1}(p_2)|D^*(q_1)\bar{D}(q_2)\rangle=&\frac{i}{\sqrt{2}}e\, f_{\chi_{c1}DD^*\gamma}\epsilon_{\gamma}^{*\mu}\epsilon_{\chi_{c1}}^{*\nu}g_{\mu\alpha}\nonumber\\
	&\times(q_{2\nu}-q_{1\nu})\epsilon_{D^*}^{\alpha},
\end{align}
\begin{align}
	\langle\gamma(p_1)\chi_{c2}(p_2)|D(q_1)\bar{D}(q_2)\rangle=e\, g_{\chi_{c2}DD\gamma}\epsilon_{\gamma}^{*\mu}\epsilon_{\chi_{c2}}^{*\alpha\beta}g_{\mu\alpha}(q_{2\beta}-q_{1\beta}),
\end{align}
\begin{align}
	\langle\gamma(p_1)\chi_{c2}(p_2)|D(q_1)\bar{D}^*(q_2)\rangle=&\;ie\, g_{\chi_{c2}DD^*\gamma}\varepsilon_{\mu\alpha\rho\beta}\epsilon_{\gamma}^{*\rho}\epsilon_{\chi_{c2}}^{*\mu\nu}\nonumber\\
	&\times q_1^{\beta}q_{2\nu}\epsilon_{\bar{D}^*}^{\alpha},
\end{align}
\begin{align}
	\langle\gamma(p_1)\chi_{c2}(p_2)|D^*(q_1)\bar{D}(q_2)\rangle=&-ie\, g_{\chi_{c2}DD^*\gamma}\varepsilon_{\mu\alpha\rho\beta}\epsilon_{\gamma}^{*\rho}\epsilon_{\chi_{c2}}^{*\mu\nu}\nonumber\\ 
	&\times q_{1\nu}q_2^{\beta}\epsilon_{D^*}^{\alpha},
\end{align}
\begin{align}
	\langle\gamma(p_1)\chi_{c2}(p_2)|D^*(q_1)\bar{D}^*(q_2)\rangle=&\;e\, g_{\chi_{c2}D^*D^*}p_2^{\lambda}\epsilon_{\gamma}^{*\rho}\epsilon_{\chi_{c2}}^{*\mu\nu}p_{1\mu}g_{\rho\nu}g_{\alpha\beta}\nonumber\\
	&\times(q_{1\lambda}-q_{2\lambda})\epsilon_{D^*}^{\alpha}\epsilon_{\bar{D}^*}^{\beta},
\end{align}
\begin{align}
	\langle D_1(q_1)\bar{D}(q_2)|\psi(p)\rangle=-g_{\psi D_1D}\epsilon_{\psi}^{\mu}\epsilon_{D_1\mu}^*,
\end{align}
\begin{align}
	\langle \gamma(p_1)D(q)|D_1(q_1)\rangle=ie\, g_{D_1D\gamma}\epsilon_{\gamma}^{*\mu}\epsilon_{D_1\mu},
\end{align}
\begin{align}
	\langle \gamma(p_1)D^*(q)|D_1(q_1)\rangle=&\;-e\, g_{D_1D^*\gamma}\varepsilon^{\mu\nu\alpha\beta}(p_{1\mu}g_{\nu\rho}-p_{1\nu}g_{\mu\rho})\nonumber\\
 &\times\epsilon_\gamma^{*\rho}\epsilon_{D^*\alpha}\epsilon_{D_1\beta}^*,
\end{align}
\begin{align}
	\langle \gamma(p_1)\chi_{c0}(p_2)|D_1(q_1)\bar{D}(q_2)\rangle=-ie\, g_{\chi_{c0}D_1D\gamma}\epsilon_{\gamma}^{*\mu}\epsilon_{D_1\mu},
\end{align}
\begin{align}
	\langle \gamma(p_1)\chi_{c0}(p_2)|D(q_1)\bar{D}_1(q_2)\rangle=ie\, g_{\chi_{c0}D_1D\gamma}\epsilon_{\gamma}^{*\mu}\epsilon_{D_1\mu},
\end{align}
\begin{align}
	\langle \gamma(p_1)\chi_{c2}(p_2)|D_1(q_1)\bar{D}(q_2)\rangle=-ie\, g_{\chi_{c2}D_1D\gamma}\epsilon_{\chi_{c2}}^{*\mu\nu}\epsilon_{\gamma\mu}^{*}\epsilon_{D_1\nu},
\end{align}
\begin{align}
	\langle \gamma(p_1)\chi_{c2}(p_2)|D(q_1)\bar{D}_1(q_2)\rangle=ie\, g_{\chi_{c2}D_1D\gamma}\epsilon_{\chi_{c2}}^{*\mu\nu}\epsilon_{\gamma\mu}^{*}\epsilon_{D_1\nu}.
\end{align}

\bibliography{gammachicj}

\begin{thebibliography}{52}%
\makeatletter
\providecommand \@ifxundefined [1]{%
 \@ifx{#1\undefined}
}%
\providecommand \@ifnum [1]{%
 \ifnum #1\expandafter \@firstoftwo
 \else \expandafter \@secondoftwo
 \fi
}%
\providecommand \@ifx [1]{%
 \ifx #1\expandafter \@firstoftwo
 \else \expandafter \@secondoftwo
 \fi
}%
\providecommand \natexlab [1]{#1}%
\providecommand \enquote  [1]{``#1''}%
\providecommand \bibnamefont  [1]{#1}%
\providecommand \bibfnamefont [1]{#1}%
\providecommand \citenamefont [1]{#1}%
\providecommand \href@noop [0]{\@secondoftwo}%
\providecommand \href [0]{\begingroup \@sanitize@url \@href}%
\providecommand \@href[1]{\@@startlink{#1}\@@href}%
\providecommand \@@href[1]{\endgroup#1\@@endlink}%
\providecommand \@sanitize@url [0]{\catcode `\\12\catcode `\$12\catcode
  `\&12\catcode `\#12\catcode `\^12\catcode `\_12\catcode `\%12\relax}%
\providecommand \@@startlink[1]{}%
\providecommand \@@endlink[0]{}%
\providecommand \url  [0]{\begingroup\@sanitize@url \@url }%
\providecommand \@url [1]{\endgroup\@href {#1}{\urlprefix }}%
\providecommand \urlprefix  [0]{URL }%
\providecommand \Eprint [0]{\href }%
\providecommand \doibase [0]{https://doi.org/}%
\providecommand \selectlanguage [0]{\@gobble}%
\providecommand \bibinfo  [0]{\@secondoftwo}%
\providecommand \bibfield  [0]{\@secondoftwo}%
\providecommand \translation [1]{[#1]}%
\providecommand \BibitemOpen [0]{}%
\providecommand \bibitemStop [0]{}%
\providecommand \bibitemNoStop [0]{.\EOS\space}%
\providecommand \EOS [0]{\spacefactor3000\relax}%
\providecommand \BibitemShut  [1]{\csname bibitem#1\endcsname}%
\let\auto@bib@innerbib\@empty
\bibitem [{\citenamefont {Choi}\ \emph {et~al.}(2003)\citenamefont {Choi} \emph
  {et~al.}}]{Belle:2003nnu}%
  \BibitemOpen
  \bibfield  {author} {\bibinfo {author} {\bibfnamefont {S.~K.}\ \bibnamefont
  {Choi}} \emph {et~al.} (\bibinfo {collaboration} {Belle}),\ }\bibfield
  {title} {\bibinfo {title} {{Observation of a narrow charmonium-like state in
  exclusive $B^\pm \to K^\pm \pi^+ \pi^- J/\psi$ decays}},\ }\href
  {https://doi.org/10.1103/PhysRevLett.91.262001} {\bibfield  {journal}
  {\bibinfo  {journal} {Phys. Rev. Lett.}\ }\textbf {\bibinfo {volume} {91}},\
  \bibinfo {pages} {262001} (\bibinfo {year} {2003})},\ \Eprint
  {https://arxiv.org/abs/hep-ex/0309032} {arXiv:hep-ex/0309032} \BibitemShut
  {NoStop}%
\bibitem [{\citenamefont {Barnes}\ \emph {et~al.}(2005)\citenamefont {Barnes},
  \citenamefont {Godfrey},\ and\ \citenamefont {Swanson}}]{Barnes:2005pb}%
  \BibitemOpen
  \bibfield  {author} {\bibinfo {author} {\bibfnamefont {T.}~\bibnamefont
  {Barnes}}, \bibinfo {author} {\bibfnamefont {S.}~\bibnamefont {Godfrey}},\
  and\ \bibinfo {author} {\bibfnamefont {E.~S.}\ \bibnamefont {Swanson}},\
  }\bibfield  {title} {\bibinfo {title} {{Higher charmonia}},\ }\href
  {https://doi.org/10.1103/PhysRevD.72.054026} {\bibfield  {journal} {\bibinfo
  {journal} {Phys. Rev. D}\ }\textbf {\bibinfo {volume} {72}},\ \bibinfo
  {pages} {054026} (\bibinfo {year} {2005})},\ \Eprint
  {https://arxiv.org/abs/hep-ph/0505002} {arXiv:hep-ph/0505002} \BibitemShut
  {NoStop}%
\bibitem [{\citenamefont {Maiani}\ \emph {et~al.}(2005)\citenamefont {Maiani},
  \citenamefont {Piccinini}, \citenamefont {Polosa},\ and\ \citenamefont
  {Riquer}}]{Maiani:2004vq}%
  \BibitemOpen
  \bibfield  {author} {\bibinfo {author} {\bibfnamefont {L.}~\bibnamefont
  {Maiani}}, \bibinfo {author} {\bibfnamefont {F.}~\bibnamefont {Piccinini}},
  \bibinfo {author} {\bibfnamefont {A.~D.}\ \bibnamefont {Polosa}},\ and\
  \bibinfo {author} {\bibfnamefont {V.}~\bibnamefont {Riquer}},\ }\bibfield
  {title} {\bibinfo {title} {{Diquark-antidiquarks with hidden or open charm
  and the nature of $X(3872)$}},\ }\href
  {https://doi.org/10.1103/PhysRevD.71.014028} {\bibfield  {journal} {\bibinfo
  {journal} {Phys. Rev. D}\ }\textbf {\bibinfo {volume} {71}},\ \bibinfo
  {pages} {014028} (\bibinfo {year} {2005})},\ \Eprint
  {https://arxiv.org/abs/hep-ph/0412098} {arXiv:hep-ph/0412098} \BibitemShut
  {NoStop}%
\bibitem [{\citenamefont {Hogaasen}\ \emph {et~al.}(2006)\citenamefont
  {Hogaasen}, \citenamefont {Richard},\ and\ \citenamefont
  {Sorba}}]{Hogaasen:2005jv}%
  \BibitemOpen
  \bibfield  {author} {\bibinfo {author} {\bibfnamefont {H.}~\bibnamefont
  {Hogaasen}}, \bibinfo {author} {\bibfnamefont {J.~M.}\ \bibnamefont
  {Richard}},\ and\ \bibinfo {author} {\bibfnamefont {P.}~\bibnamefont
  {Sorba}},\ }\bibfield  {title} {\bibinfo {title} {{A Chromomagnetic mechanism
  for the $X(3872)$ resonance}},\ }\href
  {https://doi.org/10.1103/PhysRevD.73.054013} {\bibfield  {journal} {\bibinfo
  {journal} {Phys. Rev. D}\ }\textbf {\bibinfo {volume} {73}},\ \bibinfo
  {pages} {054013} (\bibinfo {year} {2006})},\ \Eprint
  {https://arxiv.org/abs/hep-ph/0511039} {arXiv:hep-ph/0511039} \BibitemShut
  {NoStop}%
\bibitem [{\citenamefont {Cui}\ \emph {et~al.}(2007)\citenamefont {Cui},
  \citenamefont {Chen}, \citenamefont {Deng},\ and\ \citenamefont
  {Zhu}}]{Cui:2006mp}%
  \BibitemOpen
  \bibfield  {author} {\bibinfo {author} {\bibfnamefont {Y.}~\bibnamefont
  {Cui}}, \bibinfo {author} {\bibfnamefont {X.-L.}\ \bibnamefont {Chen}},
  \bibinfo {author} {\bibfnamefont {W.-Z.}\ \bibnamefont {Deng}},\ and\
  \bibinfo {author} {\bibfnamefont {S.-L.}\ \bibnamefont {Zhu}},\ }\bibfield
  {title} {\bibinfo {title} {{The Possible Heavy Tetraquarks
  $qQ\bar{q}\bar{Q}$, $qq\bar{Q}\bar{Q}$ and $qQ\bar{Q}\bar{Q}$}},\ }\href@noop
  {} {\bibfield  {journal} {\bibinfo  {journal} {HEPNP}\ }\textbf {\bibinfo
  {volume} {31}},\ \bibinfo {pages} {7} (\bibinfo {year} {2007})},\ \Eprint
  {https://arxiv.org/abs/hep-ph/0607226} {arXiv:hep-ph/0607226} \BibitemShut
  {NoStop}%
\bibitem [{\citenamefont {Wong}(2004)}]{Wong:2003xk}%
  \BibitemOpen
  \bibfield  {author} {\bibinfo {author} {\bibfnamefont {C.-Y.}\ \bibnamefont
  {Wong}},\ }\bibfield  {title} {\bibinfo {title} {{Molecular states of heavy
  quark mesons}},\ }\href {https://doi.org/10.1103/PhysRevC.69.055202}
  {\bibfield  {journal} {\bibinfo  {journal} {Phys. Rev. C}\ }\textbf {\bibinfo
  {volume} {69}},\ \bibinfo {pages} {055202} (\bibinfo {year} {2004})},\
  \Eprint {https://arxiv.org/abs/hep-ph/0311088} {arXiv:hep-ph/0311088}
  \BibitemShut {NoStop}%
\bibitem [{\citenamefont {AlFiky}\ \emph {et~al.}(2006)\citenamefont {AlFiky},
  \citenamefont {Gabbiani},\ and\ \citenamefont {Petrov}}]{AlFiky:2005jd}%
  \BibitemOpen
  \bibfield  {author} {\bibinfo {author} {\bibfnamefont {M.~T.}\ \bibnamefont
  {AlFiky}}, \bibinfo {author} {\bibfnamefont {F.}~\bibnamefont {Gabbiani}},\
  and\ \bibinfo {author} {\bibfnamefont {A.~A.}\ \bibnamefont {Petrov}},\
  }\bibfield  {title} {\bibinfo {title} {{$X(3872)$: Hadronic molecules in
  effective field theory}},\ }\href
  {https://doi.org/10.1016/j.physletb.2006.07.069} {\bibfield  {journal}
  {\bibinfo  {journal} {Phys. Lett. B}\ }\textbf {\bibinfo {volume} {640}},\
  \bibinfo {pages} {238} (\bibinfo {year} {2006})},\ \Eprint
  {https://arxiv.org/abs/hep-ph/0506141} {arXiv:hep-ph/0506141} \BibitemShut
  {NoStop}%
\bibitem [{\citenamefont {Wang}\ \emph {et~al.}(2019)\citenamefont {Wang},
  \citenamefont {Chen}, \citenamefont {Liu},\ and\ \citenamefont
  {Matsuki}}]{Wang:2019mhs}%
  \BibitemOpen
  \bibfield  {author} {\bibinfo {author} {\bibfnamefont {J.-Z.}\ \bibnamefont
  {Wang}}, \bibinfo {author} {\bibfnamefont {D.-Y.}\ \bibnamefont {Chen}},
  \bibinfo {author} {\bibfnamefont {X.}~\bibnamefont {Liu}},\ and\ \bibinfo
  {author} {\bibfnamefont {T.}~\bibnamefont {Matsuki}},\ }\bibfield  {title}
  {\bibinfo {title} {{Constructing $J/\psi$ family with updated data of
  charmoniumlike $Y$ states}},\ }\href
  {https://doi.org/10.1103/PhysRevD.99.114003} {\bibfield  {journal} {\bibinfo
  {journal} {Phys. Rev. D}\ }\textbf {\bibinfo {volume} {99}},\ \bibinfo
  {pages} {114003} (\bibinfo {year} {2019})},\ \Eprint
  {https://arxiv.org/abs/1903.07115} {arXiv:1903.07115 [hep-ph]} \BibitemShut
  {NoStop}%
\bibitem [{\citenamefont {Uehara}\ \emph {et~al.}(2006)\citenamefont {Uehara}
  \emph {et~al.}}]{Belle:2005rte}%
  \BibitemOpen
  \bibfield  {author} {\bibinfo {author} {\bibfnamefont {S.}~\bibnamefont
  {Uehara}} \emph {et~al.} (\bibinfo {collaboration} {Belle}),\ }\bibfield
  {title} {\bibinfo {title} {{Observation of a $\chi^{\prime}_{c2}$ candidate
  in $\gamma\gamma\to D\bar{D}$ production at BELLE}},\ }\href
  {https://doi.org/10.1103/PhysRevLett.96.082003} {\bibfield  {journal}
  {\bibinfo  {journal} {Phys. Rev. Lett.}\ }\textbf {\bibinfo {volume} {96}},\
  \bibinfo {pages} {082003} (\bibinfo {year} {2006})},\ \Eprint
  {https://arxiv.org/abs/hep-ex/0512035} {arXiv:hep-ex/0512035} \BibitemShut
  {NoStop}%
\bibitem [{\citenamefont {Liu}\ \emph {et~al.}(2010)\citenamefont {Liu},
  \citenamefont {Luo},\ and\ \citenamefont {Sun}}]{Liu:2009fe}%
  \BibitemOpen
  \bibfield  {author} {\bibinfo {author} {\bibfnamefont {X.}~\bibnamefont
  {Liu}}, \bibinfo {author} {\bibfnamefont {Z.-G.}\ \bibnamefont {Luo}},\ and\
  \bibinfo {author} {\bibfnamefont {Z.-F.}\ \bibnamefont {Sun}},\ }\bibfield
  {title} {\bibinfo {title} {{$X$(3915) and $X$(4350) as new members in P-wave
  charmonium family}},\ }\href {https://doi.org/10.1103/PhysRevLett.104.122001}
  {\bibfield  {journal} {\bibinfo  {journal} {Phys. Rev. Lett.}\ }\textbf
  {\bibinfo {volume} {104}},\ \bibinfo {pages} {122001} (\bibinfo {year}
  {2010})},\ \Eprint {https://arxiv.org/abs/0911.3694} {arXiv:0911.3694
  [hep-ph]} \BibitemShut {NoStop}%
\bibitem [{\citenamefont {Beringer}\ \emph {et~al.}(2012)\citenamefont
  {Beringer} \emph {et~al.}}]{ParticleDataGroup:2012pjm}%
  \BibitemOpen
  \bibfield  {author} {\bibinfo {author} {\bibfnamefont {J.}~\bibnamefont
  {Beringer}} \emph {et~al.} (\bibinfo {collaboration} {Particle Data Group}),\
  }\bibfield  {title} {\bibinfo {title} {{Review of Particle Physics (RPP)}},\
  }\href {https://doi.org/10.1103/PhysRevD.86.010001} {\bibfield  {journal}
  {\bibinfo  {journal} {Phys. Rev. D}\ }\textbf {\bibinfo {volume} {86}},\
  \bibinfo {pages} {010001} (\bibinfo {year} {2012})}\BibitemShut {NoStop}%
\bibitem [{\citenamefont {Guo}\ and\ \citenamefont
  {Meissner}(2012)}]{Guo:2012tv}%
  \BibitemOpen
  \bibfield  {author} {\bibinfo {author} {\bibfnamefont {F.-K.}\ \bibnamefont
  {Guo}}\ and\ \bibinfo {author} {\bibfnamefont {U.-G.}\ \bibnamefont
  {Meissner}},\ }\bibfield  {title} {\bibinfo {title} {{Where is the
  $\chi_{c0}(2P)$?}},\ }\href {https://doi.org/10.1103/PhysRevD.86.091501}
  {\bibfield  {journal} {\bibinfo  {journal} {Phys. Rev. D}\ }\textbf {\bibinfo
  {volume} {86}},\ \bibinfo {pages} {091501} (\bibinfo {year} {2012})},\
  \Eprint {https://arxiv.org/abs/1208.1134} {arXiv:1208.1134 [hep-ph]}
  \BibitemShut {NoStop}%
\bibitem [{\citenamefont {Olsen}(2015)}]{Olsen:2014maa}%
  \BibitemOpen
  \bibfield  {author} {\bibinfo {author} {\bibfnamefont {S.~L.}\ \bibnamefont
  {Olsen}},\ }\bibfield  {title} {\bibinfo {title} {{Is the $X$(3915) the
  $\chi_{c0}(2P)$?}},\ }\href {https://doi.org/10.1103/PhysRevD.91.057501}
  {\bibfield  {journal} {\bibinfo  {journal} {Phys. Rev. D}\ }\textbf {\bibinfo
  {volume} {91}},\ \bibinfo {pages} {057501} (\bibinfo {year} {2015})},\
  \Eprint {https://arxiv.org/abs/1410.6534} {arXiv:1410.6534 [hep-ex]}
  \BibitemShut {NoStop}%
\bibitem [{\citenamefont {Chen}\ \emph {et~al.}(2012)\citenamefont {Chen},
  \citenamefont {He}, \citenamefont {Liu},\ and\ \citenamefont
  {Matsuki}}]{Chen:2012wy}%
  \BibitemOpen
  \bibfield  {author} {\bibinfo {author} {\bibfnamefont {D.-Y.}\ \bibnamefont
  {Chen}}, \bibinfo {author} {\bibfnamefont {J.}~\bibnamefont {He}}, \bibinfo
  {author} {\bibfnamefont {X.}~\bibnamefont {Liu}},\ and\ \bibinfo {author}
  {\bibfnamefont {T.}~\bibnamefont {Matsuki}},\ }\bibfield  {title} {\bibinfo
  {title} {{Does the enhancement observed in $\gamma\gamma\to D\bar{D}$ contain
  two $P$-wave higher charmonia?}},\ }\href
  {https://doi.org/10.1140/epjc/s10052-012-2226-4} {\bibfield  {journal}
  {\bibinfo  {journal} {Eur. Phys. J. C}\ }\textbf {\bibinfo {volume} {72}},\
  \bibinfo {pages} {2226} (\bibinfo {year} {2012})},\ \Eprint
  {https://arxiv.org/abs/1207.3561} {arXiv:1207.3561 [hep-ph]} \BibitemShut
  {NoStop}%
\bibitem [{\citenamefont {Duan}\ \emph {et~al.}(2020)\citenamefont {Duan},
  \citenamefont {Luo}, \citenamefont {Liu},\ and\ \citenamefont
  {Matsuki}}]{Duan:2020tsx}%
  \BibitemOpen
  \bibfield  {author} {\bibinfo {author} {\bibfnamefont {M.-X.}\ \bibnamefont
  {Duan}}, \bibinfo {author} {\bibfnamefont {S.-Q.}\ \bibnamefont {Luo}},
  \bibinfo {author} {\bibfnamefont {X.}~\bibnamefont {Liu}},\ and\ \bibinfo
  {author} {\bibfnamefont {T.}~\bibnamefont {Matsuki}},\ }\bibfield  {title}
  {\bibinfo {title} {{Possibility of charmoniumlike state $X(3915)$ as
  $\chi_{c0}(2P)$ state}},\ }\href
  {https://doi.org/10.1103/PhysRevD.101.054029} {\bibfield  {journal} {\bibinfo
   {journal} {Phys. Rev. D}\ }\textbf {\bibinfo {volume} {101}},\ \bibinfo
  {pages} {054029} (\bibinfo {year} {2020})},\ \Eprint
  {https://arxiv.org/abs/2002.03311} {arXiv:2002.03311 [hep-ph]} \BibitemShut
  {NoStop}%
\bibitem [{\citenamefont {Aaij}\ \emph
  {et~al.}(2020{\natexlab{a}})\citenamefont {Aaij} \emph
  {et~al.}}]{LHCb:2020pxc}%
  \BibitemOpen
  \bibfield  {author} {\bibinfo {author} {\bibfnamefont {R.}~\bibnamefont
  {Aaij}} \emph {et~al.} (\bibinfo {collaboration} {LHCb}),\ }\bibfield
  {title} {\bibinfo {title} {{Amplitude analysis of the $B^+\to D^+D^-K^+$
  decay}},\ }\href {https://doi.org/10.1103/PhysRevD.102.112003} {\bibfield
  {journal} {\bibinfo  {journal} {Phys. Rev. D}\ }\textbf {\bibinfo {volume}
  {102}},\ \bibinfo {pages} {112003} (\bibinfo {year} {2020}{\natexlab{a}})},\
  \Eprint {https://arxiv.org/abs/2009.00026} {arXiv:2009.00026 [hep-ex]}
  \BibitemShut {NoStop}%
\bibitem [{\citenamefont {Aaij}\ \emph
  {et~al.}(2020{\natexlab{b}})\citenamefont {Aaij} \emph
  {et~al.}}]{LHCb:2020bls}%
  \BibitemOpen
  \bibfield  {author} {\bibinfo {author} {\bibfnamefont {R.}~\bibnamefont
  {Aaij}} \emph {et~al.} (\bibinfo {collaboration} {LHCb}),\ }\bibfield
  {title} {\bibinfo {title} {{A model-independent study of resonant structure
  in $B^+\to D^+D^-K^+$ decays}},\ }\href
  {https://doi.org/10.1103/PhysRevLett.125.242001} {\bibfield  {journal}
  {\bibinfo  {journal} {Phys. Rev. Lett.}\ }\textbf {\bibinfo {volume} {125}},\
  \bibinfo {pages} {242001} (\bibinfo {year} {2020}{\natexlab{b}})},\ \Eprint
  {https://arxiv.org/abs/2009.00025} {arXiv:2009.00025 [hep-ex]} \BibitemShut
  {NoStop}%
\bibitem [{\citenamefont {Uehara}\ \emph {et~al.}(2010)\citenamefont {Uehara}
  \emph {et~al.}}]{Belle:2009and}%
  \BibitemOpen
  \bibfield  {author} {\bibinfo {author} {\bibfnamefont {S.}~\bibnamefont
  {Uehara}} \emph {et~al.} (\bibinfo {collaboration} {Belle}),\ }\bibfield
  {title} {\bibinfo {title} {{Observation of a charmonium-like enhancement in
  the $\gamma \gamma \to \omega J/\psi$ process}},\ }\href
  {https://doi.org/10.1103/PhysRevLett.104.092001} {\bibfield  {journal}
  {\bibinfo  {journal} {Phys. Rev. Lett.}\ }\textbf {\bibinfo {volume} {104}},\
  \bibinfo {pages} {092001} (\bibinfo {year} {2010})},\ \Eprint
  {https://arxiv.org/abs/0912.4451} {arXiv:0912.4451 [hep-ex]} \BibitemShut
  {NoStop}%
\bibitem [{\citenamefont {Chen}\ \emph {et~al.}(2015)\citenamefont {Chen},
  \citenamefont {Liu},\ and\ \citenamefont {Matsuki}}]{Chen:2013yxa}%
  \BibitemOpen
  \bibfield  {author} {\bibinfo {author} {\bibfnamefont {D.-Y.}\ \bibnamefont
  {Chen}}, \bibinfo {author} {\bibfnamefont {X.}~\bibnamefont {Liu}},\ and\
  \bibinfo {author} {\bibfnamefont {T.}~\bibnamefont {Matsuki}},\ }\bibfield
  {title} {\bibinfo {title} {{Hidden-charm decays of $X(3915)$ and $Z(3930)$ as
  the $P$-wave charmonia}},\ }\href {https://doi.org/10.1093/ptep/ptv038}
  {\bibfield  {journal} {\bibinfo  {journal} {PTEP}\ }\textbf {\bibinfo
  {volume} {2015}},\ \bibinfo {pages} {043B05} (\bibinfo {year} {2015})},\
  \Eprint {https://arxiv.org/abs/1311.6274} {arXiv:1311.6274 [hep-ph]}
  \BibitemShut {NoStop}%
\bibitem [{\citenamefont {Duan}\ \emph {et~al.}(2021)\citenamefont {Duan},
  \citenamefont {Wang}, \citenamefont {Li},\ and\ \citenamefont
  {Liu}}]{Duan:2021bna}%
  \BibitemOpen
  \bibfield  {author} {\bibinfo {author} {\bibfnamefont {M.-X.}\ \bibnamefont
  {Duan}}, \bibinfo {author} {\bibfnamefont {J.-Z.}\ \bibnamefont {Wang}},
  \bibinfo {author} {\bibfnamefont {Y.-S.}\ \bibnamefont {Li}},\ and\ \bibinfo
  {author} {\bibfnamefont {X.}~\bibnamefont {Liu}},\ }\bibfield  {title}
  {\bibinfo {title} {{Role of the newly measured $ B\to KD\bar{D}$ process to
  establish $\ensuremath{\chi}_{c0}(2P)$ state}},\ }\href
  {https://doi.org/10.1103/PhysRevD.104.034035} {\bibfield  {journal} {\bibinfo
   {journal} {Phys. Rev. D}\ }\textbf {\bibinfo {volume} {104}},\ \bibinfo
  {pages} {034035} (\bibinfo {year} {2021})},\ \Eprint
  {https://arxiv.org/abs/2104.09132} {arXiv:2104.09132 [hep-ph]} \BibitemShut
  {NoStop}%
\bibitem [{\citenamefont {Ji}\ \emph {et~al.}(2023)\citenamefont {Ji},
  \citenamefont {Dong}, \citenamefont {Albaladejo}, \citenamefont {Du},
  \citenamefont {Guo}, \citenamefont {Nieves},\ and\ \citenamefont
  {Zou}}]{Ji:2022vdj}%
  \BibitemOpen
  \bibfield  {author} {\bibinfo {author} {\bibfnamefont {T.}~\bibnamefont
  {Ji}}, \bibinfo {author} {\bibfnamefont {X.-K.}\ \bibnamefont {Dong}},
  \bibinfo {author} {\bibfnamefont {M.}~\bibnamefont {Albaladejo}}, \bibinfo
  {author} {\bibfnamefont {M.-L.}\ \bibnamefont {Du}}, \bibinfo {author}
  {\bibfnamefont {F.-K.}\ \bibnamefont {Guo}}, \bibinfo {author} {\bibfnamefont
  {J.}~\bibnamefont {Nieves}},\ and\ \bibinfo {author} {\bibfnamefont {B.-S.}\
  \bibnamefont {Zou}},\ }\bibfield  {title} {\bibinfo {title} {{Understanding
  the $0^{++}$ and $2^{++}$ charmonium(-like) states near 3.9 GeV}},\ }\href
  {https://doi.org/10.1016/j.scib.2023.02.034} {\bibfield  {journal} {\bibinfo
  {journal} {Sci. Bull.}\ }\textbf {\bibinfo {volume} {68}},\ \bibinfo {pages}
  {688} (\bibinfo {year} {2023})},\ \Eprint {https://arxiv.org/abs/2212.00631}
  {arXiv:2212.00631 [hep-ph]} \BibitemShut {NoStop}%
\bibitem [{\citenamefont {Ablikim}\ \emph {et~al.}(2014)\citenamefont {Ablikim}
  \emph {et~al.}}]{BESIII:2013fnz}%
  \BibitemOpen
  \bibfield  {author} {\bibinfo {author} {\bibfnamefont {M.}~\bibnamefont
  {Ablikim}} \emph {et~al.} (\bibinfo {collaboration} {BESIII}),\ }\bibfield
  {title} {\bibinfo {title} {{Observation of $e^+e^- \to \gamma X(3872)$ at
  BESIII}},\ }\href {https://doi.org/10.1103/PhysRevLett.112.092001} {\bibfield
   {journal} {\bibinfo  {journal} {Phys. Rev. Lett.}\ }\textbf {\bibinfo
  {volume} {112}},\ \bibinfo {pages} {092001} (\bibinfo {year} {2014})},\
  \Eprint {https://arxiv.org/abs/1310.4101} {arXiv:1310.4101 [hep-ex]}
  \BibitemShut {NoStop}%
\bibitem [{\citenamefont {Ablikim}\ \emph
  {et~al.}(2019{\natexlab{a}})\citenamefont {Ablikim} \emph
  {et~al.}}]{BESIII:2019qvy}%
  \BibitemOpen
  \bibfield  {author} {\bibinfo {author} {\bibfnamefont {M.}~\bibnamefont
  {Ablikim}} \emph {et~al.} (\bibinfo {collaboration} {BESIII}),\ }\bibfield
  {title} {\bibinfo {title} {{Study of $e^+e^- \to \gamma \omega J/\psi$ and
  Observation of $X(3872) \to \omega J/\psi$}},\ }\href
  {https://doi.org/10.1103/PhysRevLett.122.232002} {\bibfield  {journal}
  {\bibinfo  {journal} {Phys. Rev. Lett.}\ }\textbf {\bibinfo {volume} {122}},\
  \bibinfo {pages} {232002} (\bibinfo {year} {2019}{\natexlab{a}})},\ \Eprint
  {https://arxiv.org/abs/1903.04695} {arXiv:1903.04695 [hep-ex]} \BibitemShut
  {NoStop}%
\bibitem [{\citenamefont {He}\ \emph {et~al.}(2014)\citenamefont {He},
  \citenamefont {Chen}, \citenamefont {Liu},\ and\ \citenamefont
  {Matsuki}}]{He:2014xna}%
  \BibitemOpen
  \bibfield  {author} {\bibinfo {author} {\bibfnamefont {L.-P.}\ \bibnamefont
  {He}}, \bibinfo {author} {\bibfnamefont {D.-Y.}\ \bibnamefont {Chen}},
  \bibinfo {author} {\bibfnamefont {X.}~\bibnamefont {Liu}},\ and\ \bibinfo
  {author} {\bibfnamefont {T.}~\bibnamefont {Matsuki}},\ }\bibfield  {title}
  {\bibinfo {title} {{Prediction of a missing higher charmonium around 4.26 GeV
  in $J/\psi$ family}},\ }\href
  {https://doi.org/10.1140/epjc/s10052-014-3208-5} {\bibfield  {journal}
  {\bibinfo  {journal} {Eur. Phys. J. C}\ }\textbf {\bibinfo {volume} {74}},\
  \bibinfo {pages} {3208} (\bibinfo {year} {2014})},\ \Eprint
  {https://arxiv.org/abs/1405.3831} {arXiv:1405.3831 [hep-ph]} \BibitemShut
  {NoStop}%
\bibitem [{\citenamefont {Li}\ and\ \citenamefont {Chao}(2009)}]{Li:2009zu}%
  \BibitemOpen
  \bibfield  {author} {\bibinfo {author} {\bibfnamefont {B.-Q.}\ \bibnamefont
  {Li}}\ and\ \bibinfo {author} {\bibfnamefont {K.-T.}\ \bibnamefont {Chao}},\
  }\bibfield  {title} {\bibinfo {title} {{Higher Charmonia and $X$, $Y$, $Z$
  states with Screened Potential}},\ }\href
  {https://doi.org/10.1103/PhysRevD.79.094004} {\bibfield  {journal} {\bibinfo
  {journal} {Phys. Rev. D}\ }\textbf {\bibinfo {volume} {79}},\ \bibinfo
  {pages} {094004} (\bibinfo {year} {2009})},\ \Eprint
  {https://arxiv.org/abs/0903.5506} {arXiv:0903.5506 [hep-ph]} \BibitemShut
  {NoStop}%
\bibitem [{\citenamefont {Chen}\ \emph {et~al.}(2018)\citenamefont {Chen},
  \citenamefont {Liu},\ and\ \citenamefont {Matsuki}}]{Chen:2017uof}%
  \BibitemOpen
  \bibfield  {author} {\bibinfo {author} {\bibfnamefont {D.-Y.}\ \bibnamefont
  {Chen}}, \bibinfo {author} {\bibfnamefont {X.}~\bibnamefont {Liu}},\ and\
  \bibinfo {author} {\bibfnamefont {T.}~\bibnamefont {Matsuki}},\ }\bibfield
  {title} {\bibinfo {title} {{Interference effect as resonance killer of newly
  observed charmoniumlike states $Y(4320)$ and $Y(4390)$}},\ }\href
  {https://doi.org/10.1140/epjc/s10052-018-5635-1} {\bibfield  {journal}
  {\bibinfo  {journal} {Eur. Phys. J. C}\ }\textbf {\bibinfo {volume} {78}},\
  \bibinfo {pages} {136} (\bibinfo {year} {2018})},\ \Eprint
  {https://arxiv.org/abs/1708.01954} {arXiv:1708.01954 [hep-ph]} \BibitemShut
  {NoStop}%
\bibitem [{\citenamefont {Guo}\ \emph {et~al.}(2013)\citenamefont {Guo},
  \citenamefont {Hanhart}, \citenamefont {Mei\ss{}ner}, \citenamefont {Wang},\
  and\ \citenamefont {Zhao}}]{Guo:2013zbw}%
  \BibitemOpen
  \bibfield  {author} {\bibinfo {author} {\bibfnamefont {F.-K.}\ \bibnamefont
  {Guo}}, \bibinfo {author} {\bibfnamefont {C.}~\bibnamefont {Hanhart}},
  \bibinfo {author} {\bibfnamefont {U.-G.}\ \bibnamefont {Mei\ss{}ner}},
  \bibinfo {author} {\bibfnamefont {Q.}~\bibnamefont {Wang}},\ and\ \bibinfo
  {author} {\bibfnamefont {Q.}~\bibnamefont {Zhao}},\ }\bibfield  {title}
  {\bibinfo {title} {{Production of the $X(3872)$ in charmonia radiative
  decays}},\ }\href {https://doi.org/10.1016/j.physletb.2013.06.053} {\bibfield
   {journal} {\bibinfo  {journal} {Phys. Lett. B}\ }\textbf {\bibinfo {volume}
  {725}},\ \bibinfo {pages} {127} (\bibinfo {year} {2013})},\ \Eprint
  {https://arxiv.org/abs/1306.3096} {arXiv:1306.3096 [hep-ph]} \BibitemShut
  {NoStop}%
\bibitem [{\citenamefont {Ding}(2009)}]{Ding:2008gr}%
  \BibitemOpen
  \bibfield  {author} {\bibinfo {author} {\bibfnamefont {G.-J.}\ \bibnamefont
  {Ding}},\ }\bibfield  {title} {\bibinfo {title} {{Are $Y(4260)$ and
  $Z^+_2(4250)$ are $D_1D$ or $D_0D^*$ Hadronic Molecules?}},\ }\href
  {https://doi.org/10.1103/PhysRevD.79.014001} {\bibfield  {journal} {\bibinfo
  {journal} {Phys. Rev. D}\ }\textbf {\bibinfo {volume} {79}},\ \bibinfo
  {pages} {014001} (\bibinfo {year} {2009})},\ \Eprint
  {https://arxiv.org/abs/0809.4818} {arXiv:0809.4818 [hep-ph]} \BibitemShut
  {NoStop}%
\bibitem [{\citenamefont {Li}\ \emph {et~al.}(2013)\citenamefont {Li},
  \citenamefont {Wang}, \citenamefont {Dong},\ and\ \citenamefont
  {Zhang}}]{Li:2013bca}%
  \BibitemOpen
  \bibfield  {author} {\bibinfo {author} {\bibfnamefont {M.-T.}\ \bibnamefont
  {Li}}, \bibinfo {author} {\bibfnamefont {W.-L.}\ \bibnamefont {Wang}},
  \bibinfo {author} {\bibfnamefont {Y.-B.}\ \bibnamefont {Dong}},\ and\
  \bibinfo {author} {\bibfnamefont {Z.-Y.}\ \bibnamefont {Zhang}},\ }\bibfield
  {title} {\bibinfo {title} {{A Study of $P$-wave Heavy Meson Interactions in A
  Chiral Quark Model}},\ }\href@noop {} {\  (\bibinfo {year} {2013})},\ \Eprint
  {https://arxiv.org/abs/1303.4140} {arXiv:1303.4140 [nucl-th]} \BibitemShut
  {NoStop}%
\bibitem [{\citenamefont {Wang}\ \emph {et~al.}(2013)\citenamefont {Wang},
  \citenamefont {Hanhart},\ and\ \citenamefont {Zhao}}]{Wang:2013cya}%
  \BibitemOpen
  \bibfield  {author} {\bibinfo {author} {\bibfnamefont {Q.}~\bibnamefont
  {Wang}}, \bibinfo {author} {\bibfnamefont {C.}~\bibnamefont {Hanhart}},\ and\
  \bibinfo {author} {\bibfnamefont {Q.}~\bibnamefont {Zhao}},\ }\bibfield
  {title} {\bibinfo {title} {{Decoding the riddle of $Y(4260)$ and
  $Z_c(3900)$}},\ }\href {https://doi.org/10.1103/PhysRevLett.111.132003}
  {\bibfield  {journal} {\bibinfo  {journal} {Phys. Rev. Lett.}\ }\textbf
  {\bibinfo {volume} {111}},\ \bibinfo {pages} {132003} (\bibinfo {year}
  {2013})},\ \Eprint {https://arxiv.org/abs/1303.6355} {arXiv:1303.6355
  [hep-ph]} \BibitemShut {NoStop}%
\bibitem [{\citenamefont {Ablikim}\ \emph {et~al.}(2017)\citenamefont {Ablikim}
  \emph {et~al.}}]{BESIII:2016bnd}%
  \BibitemOpen
  \bibfield  {author} {\bibinfo {author} {\bibfnamefont {M.}~\bibnamefont
  {Ablikim}} \emph {et~al.} (\bibinfo {collaboration} {BESIII}),\ }\bibfield
  {title} {\bibinfo {title} {{Precise measurement of the $e^+e^-\to
  \pi^+\pi^-J/\psi$ cross section at center-of-mass energies from 3.77 to 4.60
  GeV}},\ }\href {https://doi.org/10.1103/PhysRevLett.118.092001} {\bibfield
  {journal} {\bibinfo  {journal} {Phys. Rev. Lett.}\ }\textbf {\bibinfo
  {volume} {118}},\ \bibinfo {pages} {092001} (\bibinfo {year} {2017})},\
  \Eprint {https://arxiv.org/abs/1611.01317} {arXiv:1611.01317 [hep-ex]}
  \BibitemShut {NoStop}%
\bibitem [{\citenamefont {Wang}\ \emph {et~al.}(2020)\citenamefont {Wang},
  \citenamefont {Qian}, \citenamefont {Liu},\ and\ \citenamefont
  {Matsuki}}]{Wang:2020prx}%
  \BibitemOpen
  \bibfield  {author} {\bibinfo {author} {\bibfnamefont {J.-Z.}\ \bibnamefont
  {Wang}}, \bibinfo {author} {\bibfnamefont {R.-Q.}\ \bibnamefont {Qian}},
  \bibinfo {author} {\bibfnamefont {X.}~\bibnamefont {Liu}},\ and\ \bibinfo
  {author} {\bibfnamefont {T.}~\bibnamefont {Matsuki}},\ }\bibfield  {title}
  {\bibinfo {title} {{Are the $Y$ states around 4.6 GeV from $e^+e^-$
  annihilation higher charmonia?}},\ }\href
  {https://doi.org/10.1103/PhysRevD.101.034001} {\bibfield  {journal} {\bibinfo
   {journal} {Phys. Rev. D}\ }\textbf {\bibinfo {volume} {101}},\ \bibinfo
  {pages} {034001} (\bibinfo {year} {2020})},\ \Eprint
  {https://arxiv.org/abs/2001.00175} {arXiv:2001.00175 [hep-ph]} \BibitemShut
  {NoStop}%
\bibitem [{\citenamefont {Lu}\ \emph {et~al.}(2017)\citenamefont {Lu},
  \citenamefont {Anwar},\ and\ \citenamefont {Zou}}]{Lu:2017yhl}%
  \BibitemOpen
  \bibfield  {author} {\bibinfo {author} {\bibfnamefont {Y.}~\bibnamefont
  {Lu}}, \bibinfo {author} {\bibfnamefont {M.~N.}\ \bibnamefont {Anwar}},\ and\
  \bibinfo {author} {\bibfnamefont {B.-S.}\ \bibnamefont {Zou}},\ }\bibfield
  {title} {\bibinfo {title} {{$X(4260)$ Revisited: A Coupled Channel
  Perspective}},\ }\href {https://doi.org/10.1103/PhysRevD.96.114022}
  {\bibfield  {journal} {\bibinfo  {journal} {Phys. Rev. D}\ }\textbf {\bibinfo
  {volume} {96}},\ \bibinfo {pages} {114022} (\bibinfo {year} {2017})},\
  \Eprint {https://arxiv.org/abs/1705.00449} {arXiv:1705.00449 [hep-ph]}
  \BibitemShut {NoStop}%
\bibitem [{\citenamefont {Man}\ \emph {et~al.}(2025)\citenamefont {Man},
  \citenamefont {Luo}, \citenamefont {Bai},\ and\ \citenamefont
  {Liu}}]{Man:2025zfu}%
  \BibitemOpen
  \bibfield  {author} {\bibinfo {author} {\bibfnamefont {Z.-L.}\ \bibnamefont
  {Man}}, \bibinfo {author} {\bibfnamefont {S.-Q.}\ \bibnamefont {Luo}},
  \bibinfo {author} {\bibfnamefont {Z.-Y.}\ \bibnamefont {Bai}},\ and\ \bibinfo
  {author} {\bibfnamefont {X.}~\bibnamefont {Liu}},\ }\bibfield  {title}
  {\bibinfo {title} {{Unraveling charmonium mixing scheme for the $\psi(4220)$
  and $\psi(4380)$ by a coupled-channel approach}},\ }\href@noop {} {\
  (\bibinfo {year} {2025})},\ \Eprint {https://arxiv.org/abs/2502.08072}
  {arXiv:2502.08072 [hep-ph]} \BibitemShut {NoStop}%
\bibitem [{\citenamefont {Ablikim}\ \emph {et~al.}(2005)\citenamefont {Ablikim}
  \emph {et~al.}}]{BES:2005bmx}%
  \BibitemOpen
  \bibfield  {author} {\bibinfo {author} {\bibfnamefont {M.}~\bibnamefont
  {Ablikim}} \emph {et~al.} (\bibinfo {collaboration} {BES}),\ }\bibfield
  {title} {\bibinfo {title} {{Precise measurement of spin-averaged
  $\chi_{cJ}(1P)$ mass using photon conversions in $\psi(2S) \to \gamma
  \chi_{cJ}$}},\ }\href {https://doi.org/10.1103/PhysRevD.71.092002} {\bibfield
   {journal} {\bibinfo  {journal} {Phys. Rev. D}\ }\textbf {\bibinfo {volume}
  {71}},\ \bibinfo {pages} {092002} (\bibinfo {year} {2005})},\ \Eprint
  {https://arxiv.org/abs/hep-ex/0502031} {arXiv:hep-ex/0502031} \BibitemShut
  {NoStop}%
\bibitem [{\citenamefont {Artuso}\ \emph {et~al.}(2005)\citenamefont {Artuso}
  \emph {et~al.}}]{CLEO:2004jkt}%
  \BibitemOpen
  \bibfield  {author} {\bibinfo {author} {\bibfnamefont {M.}~\bibnamefont
  {Artuso}} \emph {et~al.} (\bibinfo {collaboration} {CLEO}),\ }\bibfield
  {title} {\bibinfo {title} {{Photon transitions in $\Upsilon(2S)$ and
  $\Upsilon(3S)$ decays}},\ }\href
  {https://doi.org/10.1103/PhysRevLett.94.032001} {\bibfield  {journal}
  {\bibinfo  {journal} {Phys. Rev. Lett.}\ }\textbf {\bibinfo {volume} {94}},\
  \bibinfo {pages} {032001} (\bibinfo {year} {2005})},\ \Eprint
  {https://arxiv.org/abs/hep-ex/0411068} {arXiv:hep-ex/0411068} \BibitemShut
  {NoStop}%
\bibitem [{\citenamefont {Liu}\ \emph {et~al.}(2006)\citenamefont {Liu},
  \citenamefont {Zeng},\ and\ \citenamefont {Li}}]{Liu:2006dq}%
  \BibitemOpen
  \bibfield  {author} {\bibinfo {author} {\bibfnamefont {X.}~\bibnamefont
  {Liu}}, \bibinfo {author} {\bibfnamefont {X.-Q.}\ \bibnamefont {Zeng}},\ and\
  \bibinfo {author} {\bibfnamefont {X.-Q.}\ \bibnamefont {Li}},\ }\bibfield
  {title} {\bibinfo {title} {{Study on contributions of hadronic loops to
  decays of $J/\psi \to vector + pseudoscalar$ mesons}},\ }\href
  {https://doi.org/10.1103/PhysRevD.74.074003} {\bibfield  {journal} {\bibinfo
  {journal} {Phys. Rev. D}\ }\textbf {\bibinfo {volume} {74}},\ \bibinfo
  {pages} {074003} (\bibinfo {year} {2006})},\ \Eprint
  {https://arxiv.org/abs/hep-ph/0606191} {arXiv:hep-ph/0606191} \BibitemShut
  {NoStop}%
\bibitem [{\citenamefont {Liu}\ \emph {et~al.}(2009)\citenamefont {Liu},
  \citenamefont {Zhang},\ and\ \citenamefont {Li}}]{Liu:2009dr}%
  \BibitemOpen
  \bibfield  {author} {\bibinfo {author} {\bibfnamefont {X.}~\bibnamefont
  {Liu}}, \bibinfo {author} {\bibfnamefont {B.}~\bibnamefont {Zhang}},\ and\
  \bibinfo {author} {\bibfnamefont {X.-Q.}\ \bibnamefont {Li}},\ }\bibfield
  {title} {\bibinfo {title} {{The puzzle of excessive non-$D\bar{D}$ component
  of the inclusive $\ensuremath{\psi}(3770)$ decay and the long-distant
  contribution}},\ }\href {https://doi.org/10.1016/j.physletb.2009.04.047}
  {\bibfield  {journal} {\bibinfo  {journal} {Phys. Lett. B}\ }\textbf
  {\bibinfo {volume} {675}},\ \bibinfo {pages} {441} (\bibinfo {year}
  {2009})},\ \Eprint {https://arxiv.org/abs/0902.0480} {arXiv:0902.0480
  [hep-ph]} \BibitemShut {NoStop}%
\bibitem [{\citenamefont {Zhang}\ \emph {et~al.}(2009)\citenamefont {Zhang},
  \citenamefont {Li},\ and\ \citenamefont {Zhao}}]{Zhang:2009kr}%
  \BibitemOpen
  \bibfield  {author} {\bibinfo {author} {\bibfnamefont {Y.-J.}\ \bibnamefont
  {Zhang}}, \bibinfo {author} {\bibfnamefont {G.}~\bibnamefont {Li}},\ and\
  \bibinfo {author} {\bibfnamefont {Q.}~\bibnamefont {Zhao}},\ }\bibfield
  {title} {\bibinfo {title} {{Towards a dynamical understanding of the
  non-$D\bar{D}$ decay of $\psi(3770)$}},\ }\href
  {https://doi.org/10.1103/PhysRevLett.102.172001} {\bibfield  {journal}
  {\bibinfo  {journal} {Phys. Rev. Lett.}\ }\textbf {\bibinfo {volume} {102}},\
  \bibinfo {pages} {172001} (\bibinfo {year} {2009})},\ \Eprint
  {https://arxiv.org/abs/0902.1300} {arXiv:0902.1300 [hep-ph]} \BibitemShut
  {NoStop}%
\bibitem [{\citenamefont {Chen}\ \emph {et~al.}(2013)\citenamefont {Chen},
  \citenamefont {Liu},\ and\ \citenamefont {Matsuki}}]{Chen:2013cpa}%
  \BibitemOpen
  \bibfield  {author} {\bibinfo {author} {\bibfnamefont {D.-Y.}\ \bibnamefont
  {Chen}}, \bibinfo {author} {\bibfnamefont {X.}~\bibnamefont {Liu}},\ and\
  \bibinfo {author} {\bibfnamefont {T.}~\bibnamefont {Matsuki}},\ }\bibfield
  {title} {\bibinfo {title} {{Anomalous radiative transitions between $h_b(nP)$
  and $\eta_b(mS)$ and hadronic loop effect}},\ }\href
  {https://doi.org/10.1103/PhysRevD.87.094010} {\bibfield  {journal} {\bibinfo
  {journal} {Phys. Rev. D}\ }\textbf {\bibinfo {volume} {87}},\ \bibinfo
  {pages} {094010} (\bibinfo {year} {2013})},\ \Eprint
  {https://arxiv.org/abs/1304.0372} {arXiv:1304.0372 [hep-ph]} \BibitemShut
  {NoStop}%
\bibitem [{\citenamefont {Cheng}\ \emph {et~al.}(2005)\citenamefont {Cheng},
  \citenamefont {Chua},\ and\ \citenamefont {Soni}}]{Cheng:2004ru}%
  \BibitemOpen
  \bibfield  {author} {\bibinfo {author} {\bibfnamefont {H.-Y.}\ \bibnamefont
  {Cheng}}, \bibinfo {author} {\bibfnamefont {C.-K.}\ \bibnamefont {Chua}},\
  and\ \bibinfo {author} {\bibfnamefont {A.}~\bibnamefont {Soni}},\ }\bibfield
  {title} {\bibinfo {title} {{Final state interactions in hadronic $B$
  decays}},\ }\href {https://doi.org/10.1103/PhysRevD.71.014030} {\bibfield
  {journal} {\bibinfo  {journal} {Phys. Rev. D}\ }\textbf {\bibinfo {volume}
  {71}},\ \bibinfo {pages} {014030} (\bibinfo {year} {2005})},\ \Eprint
  {https://arxiv.org/abs/hep-ph/0409317} {arXiv:hep-ph/0409317} \BibitemShut
  {NoStop}%
\bibitem [{\citenamefont {Duan}(2024)}]{Duan:2024zuo}%
  \BibitemOpen
  \bibfield  {author} {\bibinfo {author} {\bibfnamefont {M.-X.}\ \bibnamefont
  {Duan}},\ }\bibfield  {title} {\bibinfo {title} {{Role of the short-distance
  interaction in $e^+e^-\to\ensuremath{\gamma}X(3872)$}},\ }\href
  {https://doi.org/10.1103/PhysRevD.110.034027} {\bibfield  {journal} {\bibinfo
   {journal} {Phys. Rev. D}\ }\textbf {\bibinfo {volume} {110}},\ \bibinfo
  {pages} {034027} (\bibinfo {year} {2024})},\ \Eprint
  {https://arxiv.org/abs/2403.06440} {arXiv:2403.06440 [hep-ph]} \BibitemShut
  {NoStop}%
\bibitem [{\citenamefont {Casalbuoni}\ \emph {et~al.}(1997)\citenamefont
  {Casalbuoni}, \citenamefont {Deandrea}, \citenamefont {Di~Bartolomeo},
  \citenamefont {Gatto}, \citenamefont {Feruglio},\ and\ \citenamefont
  {Nardulli}}]{Casalbuoni:1996pg}%
  \BibitemOpen
  \bibfield  {author} {\bibinfo {author} {\bibfnamefont {R.}~\bibnamefont
  {Casalbuoni}}, \bibinfo {author} {\bibfnamefont {A.}~\bibnamefont
  {Deandrea}}, \bibinfo {author} {\bibfnamefont {N.}~\bibnamefont
  {Di~Bartolomeo}}, \bibinfo {author} {\bibfnamefont {R.}~\bibnamefont
  {Gatto}}, \bibinfo {author} {\bibfnamefont {F.}~\bibnamefont {Feruglio}},\
  and\ \bibinfo {author} {\bibfnamefont {G.}~\bibnamefont {Nardulli}},\
  }\bibfield  {title} {\bibinfo {title} {{Phenomenology of heavy meson chiral
  Lagrangians}},\ }\href {https://doi.org/10.1016/S0370-1573(96)00027-0}
  {\bibfield  {journal} {\bibinfo  {journal} {Phys. Rept.}\ }\textbf {\bibinfo
  {volume} {281}},\ \bibinfo {pages} {145} (\bibinfo {year} {1997})},\ \Eprint
  {https://arxiv.org/abs/hep-ph/9605342} {arXiv:hep-ph/9605342} \BibitemShut
  {NoStop}%
\bibitem [{\citenamefont {Colangelo}\ \emph {et~al.}(2004)\citenamefont
  {Colangelo}, \citenamefont {De~Fazio},\ and\ \citenamefont
  {Pham}}]{Colangelo:2003sa}%
  \BibitemOpen
  \bibfield  {author} {\bibinfo {author} {\bibfnamefont {P.}~\bibnamefont
  {Colangelo}}, \bibinfo {author} {\bibfnamefont {F.}~\bibnamefont
  {De~Fazio}},\ and\ \bibinfo {author} {\bibfnamefont {T.~N.}\ \bibnamefont
  {Pham}},\ }\bibfield  {title} {\bibinfo {title} {{Nonfactorizable
  contributions in B decays to charmonium: The Case of $B^- \to K^- h_c$}},\
  }\href {https://doi.org/10.1103/PhysRevD.69.054023} {\bibfield  {journal}
  {\bibinfo  {journal} {Phys. Rev. D}\ }\textbf {\bibinfo {volume} {69}},\
  \bibinfo {pages} {054023} (\bibinfo {year} {2004})},\ \Eprint
  {https://arxiv.org/abs/hep-ph/0310084} {arXiv:hep-ph/0310084} \BibitemShut
  {NoStop}%
\bibitem [{\citenamefont {Xu}\ \emph {et~al.}(2016)\citenamefont {Xu},
  \citenamefont {Liu},\ and\ \citenamefont {Matsuki}}]{Xu:2016kbn}%
  \BibitemOpen
  \bibfield  {author} {\bibinfo {author} {\bibfnamefont {H.}~\bibnamefont
  {Xu}}, \bibinfo {author} {\bibfnamefont {X.}~\bibnamefont {Liu}},\ and\
  \bibinfo {author} {\bibfnamefont {T.}~\bibnamefont {Matsuki}},\ }\bibfield
  {title} {\bibinfo {title} {{Understanding $B^-\to X(3823)K^-$ via
  rescattering mechanism and predicting $B^-\to \eta_{c2}
  (^1D_2)/\psi_3(^3D_3)K^-$}},\ }\href
  {https://doi.org/10.1103/PhysRevD.94.034005} {\bibfield  {journal} {\bibinfo
  {journal} {Phys. Rev. D}\ }\textbf {\bibinfo {volume} {94}},\ \bibinfo
  {pages} {034005} (\bibinfo {year} {2016})},\ \Eprint
  {https://arxiv.org/abs/1605.04776} {arXiv:1605.04776 [hep-ph]} \BibitemShut
  {NoStop}%
\bibitem [{\citenamefont {Wang}\ and\ \citenamefont
  {Liu}(2024)}]{Wang:2023zxj}%
  \BibitemOpen
  \bibfield  {author} {\bibinfo {author} {\bibfnamefont {J.-Z.}\ \bibnamefont
  {Wang}}\ and\ \bibinfo {author} {\bibfnamefont {X.}~\bibnamefont {Liu}},\
  }\bibfield  {title} {\bibinfo {title} {{Identifying a characterized energy
  level structure of higher charmonium well matched to the peak structures in
  $e^+e^-\to\pi^+D^0D^{*-}$}},\ }\href
  {https://doi.org/10.1016/j.physletb.2024.138456} {\bibfield  {journal}
  {\bibinfo  {journal} {Phys. Lett. B}\ }\textbf {\bibinfo {volume} {849}},\
  \bibinfo {pages} {138456} (\bibinfo {year} {2024})},\ \Eprint
  {https://arxiv.org/abs/2306.14695} {arXiv:2306.14695 [hep-ph]} \BibitemShut
  {NoStop}%
\bibitem [{\citenamefont {Ablikim}\ \emph
  {et~al.}(2019{\natexlab{b}})\citenamefont {Ablikim} \emph
  {et~al.}}]{BESIII:2018iea}%
  \BibitemOpen
  \bibfield  {author} {\bibinfo {author} {\bibfnamefont {M.}~\bibnamefont
  {Ablikim}} \emph {et~al.} (\bibinfo {collaboration} {BESIII}),\ }\bibfield
  {title} {\bibinfo {title} {{Evidence of a resonant structure in the
  $e^+e^-\to \pi^+D^0D^{*-}$ cross section between 4.05 and 4.60 GeV}},\ }\href
  {https://doi.org/10.1103/PhysRevLett.122.102002} {\bibfield  {journal}
  {\bibinfo  {journal} {Phys. Rev. Lett.}\ }\textbf {\bibinfo {volume} {122}},\
  \bibinfo {pages} {102002} (\bibinfo {year} {2019}{\natexlab{b}})},\ \Eprint
  {https://arxiv.org/abs/1808.02847} {arXiv:1808.02847 [hep-ex]} \BibitemShut
  {NoStop}%
\bibitem [{\citenamefont {Liu}\ \emph {et~al.}(2024)\citenamefont {Liu},
  \citenamefont {Ling},\ and\ \citenamefont {Geng}}]{Liu:2024ziu}%
  \BibitemOpen
  \bibfield  {author} {\bibinfo {author} {\bibfnamefont {M.-Z.}\ \bibnamefont
  {Liu}}, \bibinfo {author} {\bibfnamefont {X.-Z.}\ \bibnamefont {Ling}},\ and\
  \bibinfo {author} {\bibfnamefont {L.-S.}\ \bibnamefont {Geng}},\ }\bibfield
  {title} {\bibinfo {title} {{Productions of X(3872)/Zc(3900) and
  X2(4013)/Zc(4020) in Y(4220) and Y(4360) decays}},\ }\href
  {https://doi.org/10.1103/PhysRevD.110.054035} {\bibfield  {journal} {\bibinfo
   {journal} {Phys. Rev. D}\ }\textbf {\bibinfo {volume} {110}},\ \bibinfo
  {pages} {054035} (\bibinfo {year} {2024})},\ \Eprint
  {https://arxiv.org/abs/2404.07681} {arXiv:2404.07681 [hep-ph]} \BibitemShut
  {NoStop}%
\bibitem [{\citenamefont {Workman}\ \emph {et~al.}(2022)\citenamefont {Workman}
  \emph {et~al.}}]{ParticleDataGroup:2022pth}%
  \BibitemOpen
  \bibfield  {author} {\bibinfo {author} {\bibfnamefont {R.~L.}\ \bibnamefont
  {Workman}} \emph {et~al.} (\bibinfo {collaboration} {Particle Data Group}),\
  }\bibfield  {title} {\bibinfo {title} {{Review of Particle Physics}},\ }\href
  {https://doi.org/10.1093/ptep/ptac097} {\bibfield  {journal} {\bibinfo
  {journal} {PTEP}\ }\textbf {\bibinfo {volume} {2022}},\ \bibinfo {pages}
  {083C01} (\bibinfo {year} {2022})}\BibitemShut {NoStop}%
\bibitem [{\citenamefont {Chen}\ \emph {et~al.}(2010)\citenamefont {Chen},
  \citenamefont {Dong},\ and\ \citenamefont {Liu}}]{Chen:2010re}%
  \BibitemOpen
  \bibfield  {author} {\bibinfo {author} {\bibfnamefont {D.-Y.}\ \bibnamefont
  {Chen}}, \bibinfo {author} {\bibfnamefont {Y.-B.}\ \bibnamefont {Dong}},\
  and\ \bibinfo {author} {\bibfnamefont {X.}~\bibnamefont {Liu}},\ }\bibfield
  {title} {\bibinfo {title} {{Long-distant contribution and $\chi_{c1}$
  radiative decays to light vector meson}},\ }\href
  {https://doi.org/10.1140/epjc/s10052-010-1449-5} {\bibfield  {journal}
  {\bibinfo  {journal} {Eur. Phys. J. C}\ }\textbf {\bibinfo {volume} {70}},\
  \bibinfo {pages} {177} (\bibinfo {year} {2010})},\ \Eprint
  {https://arxiv.org/abs/1005.0066} {arXiv:1005.0066 [hep-ph]} \BibitemShut
  {NoStop}%
\bibitem [{\citenamefont {Godfrey}\ and\ \citenamefont
  {Moats}(2016)}]{Godfrey:2015dva}%
  \BibitemOpen
  \bibfield  {author} {\bibinfo {author} {\bibfnamefont {S.}~\bibnamefont
  {Godfrey}}\ and\ \bibinfo {author} {\bibfnamefont {K.}~\bibnamefont
  {Moats}},\ }\bibfield  {title} {\bibinfo {title} {{Properties of Excited
  Charm and Charm-Strange Mesons}},\ }\href
  {https://doi.org/10.1103/PhysRevD.93.034035} {\bibfield  {journal} {\bibinfo
  {journal} {Phys. Rev. D}\ }\textbf {\bibinfo {volume} {93}},\ \bibinfo
  {pages} {034035} (\bibinfo {year} {2016})},\ \Eprint
  {https://arxiv.org/abs/1510.08305} {arXiv:1510.08305 [hep-ph]} \BibitemShut
  {NoStop}%
\bibitem [{\citenamefont {Li}\ \emph {et~al.}(2021)\citenamefont {Li},
  \citenamefont {Bai}, \citenamefont {Huang},\ and\ \citenamefont
  {Liu}}]{Li:2021jjt}%
  \BibitemOpen
  \bibfield  {author} {\bibinfo {author} {\bibfnamefont {Y.-S.}\ \bibnamefont
  {Li}}, \bibinfo {author} {\bibfnamefont {Z.-Y.}\ \bibnamefont {Bai}},
  \bibinfo {author} {\bibfnamefont {Q.}~\bibnamefont {Huang}},\ and\ \bibinfo
  {author} {\bibfnamefont {X.}~\bibnamefont {Liu}},\ }\bibfield  {title}
  {\bibinfo {title} {{Hidden-bottom hadronic decays of $\Upsilon{}(10753)$ with
  a $\ensuremath{\eta}^{(')}$ or $\ensuremath{\omega}$ emission}},\ }\href
  {https://doi.org/10.1103/PhysRevD.104.034036} {\bibfield  {journal} {\bibinfo
   {journal} {Phys. Rev. D}\ }\textbf {\bibinfo {volume} {104}},\ \bibinfo
  {pages} {034036} (\bibinfo {year} {2021})},\ \Eprint
  {https://arxiv.org/abs/2106.14123} {arXiv:2106.14123 [hep-ph]} \BibitemShut
  {NoStop}%
\end{thebibliography}%

\end{document}